\documentclass[manuscript, acmsmall, xcolor=svgnames,usenames,dvipsnames]{acmart}
\usepackage{svg}
\usepackage{algorithm}

\setcopyright{none} 
\acmDOI{} 
\acmISBN{}

\usepackage[figuresright]{rotating} \usepackage{graphicx}
 \usepackage{booktabs}
 \usepackage{array}

\usepackage{algpseudocode}
\usepackage{tcolorbox}
\usepackage{multirow}

\usepackage{xcolor}
\usepackage{float}
\usepackage{pgfplots}
\usepgfplotslibrary{groupplots}
\def\norminf#1{\|#1\|_\infty}

\usepackage{listings}
\definecolor{gray}{rgb}{0.5,0.5,0.5}
\definecolor{mauve}{rgb}{0.58,0,0.82}
\definecolor{lightgrey}{rgb}{0.9,0.9,0.9}
\definecolor{darkgreen}{rgb}{0,0.6,0}

\newcommand{\ve}[1]{{\mathbf{#1}}}

\newcommand{\ive}[1]{{\boldsymbol{#1}}}

\newcommand{\Nes}[0]{{N_{\mathrm{ens}}}}
\newcommand{\nfma}[0]{{N_{\mathrm{FMA}}}}

\newcommand{\neab}[0]{{n_{\mathrm{eab}}}}

\newcommand{\fl}[0]{{{\mathrm{fl}}}}
\newcommand{\psumsig}[0]{{S_{p_i,\mathrm{sum}}}}
\newcommand{\psumsigd}[0]{{S'_{p_i,\mathrm{sum}}}}
\newcommand{\Sacc}[0]{S_{\mathrm{acc}}}

\definecolor{matlabblue}{RGB}{0,0,255}        
\definecolor{matlabgreen}{RGB}{34,139,34}     
\definecolor{matlabpurple}{RGB}{160,32,240}   
\definecolor{matlabgray}{RGB}{128,128,128}    
\definecolor{matlabbg}{RGB}{255,255,255}      
\definecolor{mypink}{RGB}{255,105,180}

\newif\ifShowCorrections
\ShowCorrectionstrue
\ifShowCorrections
\usepackage[normalem]{ulem}
\definecolor{forgreen}{rgb}{0,0.6,0}
\definecolor{orange}{RGB}{255,140,0}
\definecolor{skyblue}{RGB}{100, 150, 235}

\newenvironment{mmc}{\begin{quote}\color{ForestGreen}\small\sf{$\clubsuit$~MM~}}{\end{quote}}

\newcommand{\sw}[2]{{\color{red}\sout{#1}}{\color{blue}#2}}
\newcommand{\swc}[1]{{\color{orange}[#1]}}
\else

\newcommand{\mmc}[1]{}
\newcommand{\sw}[2]{#2}
\newcommand{\swc}[1]{}
\fi

\AtBeginDocument{%
  }

\makeatletter \def\@formatdoi#1{} \makeatother
\makeatletter \renewcommand\footnotetextcopyrightpermission[1]{} \makeatother

\AtBeginDocument{ \fancypagestyle{firstpagestyle}{ \fancyhf{} } }

\makeatletter \AtBeginDocument{ \fancyfoot[RO,LE]{} } \makeatother

\begin{document}

\title{Accurate Models of AMD Matrix Cores}

\author{Faizan A. Khattak}
\email{f.a.khattak@leeds.ac.uk}
\affiliation{%
  \institution{University of Leeds, Leeds}
  \city{Leeds}
  \country{UK}
  }
\author{Mantas Mikaitis}
\email{mmikaitis@leeds.ac.uk}
\affiliation{%
  \institution{University of Leeds, Leeds}
  \city{Leeds}
  \country{UK}
}
\author{Carlo J. Graziani}
\email{cgraziani@anl.gov}
\affiliation{%
  \institution{Argonne National Laboratory}
  \city{Lemont, IL}
  \country{USA}
}

\renewcommand{\shortauthors}{Khattak, Mikaitis, and Graziani}

\begin{abstract}
  Matrix multipliers available on recent GPUs do not conform with the IEEE 754 floating point standard.
  Features of matrix multipliers differ across vendors and architectures of the same vendor, such as accumulator width, rounding behaviour, normalisation points, intermediate underflow and overflow logic, the handling of subnormals, and the treatment of special inputs.
  As a result, reproducibility of small matrix multiplier results across devices is not possible and cannot be achieved by software control.
  Implementation details of matrix multipliers are not documented, making it difficult to interpret discrepancies in the computed results.
We characterise the numerical behaviour of matrix multipliers across three AMD GPU architectures: CDNA 1, CDNA 2, and CDNA 3, using the MI100, MI210/250, and MI300A/300X GPUs, respectively.
We design test vectors to target numerical features for all supported input formats and provide the derivation and the reasoning for why each vector allows to determine a particular numerical feature based on the outputs of the devices.
MATLAB-based software models of the matrix multipliers are then developed for each architecture and validated for bit-level reproducibility against hardware using a randomized test suite consisting of 10 million sets of random input vectors.
To achieve this, we applied a previously developed technique to iteratively refine the accuracy of the models in a loop, by randomized testing followed by test-refinement until the model matches the hardware for every test case.
Finally, as a proof of concept for what experimental research can be done with the models, we have utilised them in two demonstrative numerical applications, quantifying application-level accuracy differences between AMD matrix cores and the NVIDIA tensor cores.
\end{abstract}



\keywords{Matrix cores, mixed-precision computing, matrix multiply, inner product, IEEE 754 standard arithmetic, OCP low-precision formats}


\maketitle

\fbox{
  \parbox{5.3in}{The software associated with this paper, the MATLAB Tensor Core v0.6, which includes all three CDNA architecture matrix core models as well as many NVIDIA tensor core models and a generalised model that can be used to instantiate custom variants of matrix multipliers, is available on GitHub: \url{https://github.com/north-numerical-computing/MATLAB-tensor-core}.}
}

\section{Introduction}

Early work by Hickmann~and~Bradford~\cite{hibr19} on the NVIDIA V100 GPU showed how to determine several numerical features of matrix multipliers that were not documented, such as the rounding mode and internal precision of the accumulator.
Subsequently, \cite{fhmp21,llfs24,vlpg25} conducted experiments on the successors to the V100, the A100, T4, and the H100, as well as GPUs from AMD, the MI100, and MI250X, demonstrating differences between the V100, A100 and H100, for example, in the accumulator's precision within the dot products, by showing that 24-bit, 25-bit and 26-bit accumulation was used, respectively.
Due to the absence of documented numerical behaviour, existing software-based matrix multiplier models may fail to accurately represent hardware implementations.
In many cases, simulations rely on IEEE-compliant models, which do not faithfully capture the behaviour of these specialised units.
Developing accurate software models is therefore essential—not only for understanding hardware behaviour but also for enabling reliable scientific computing and ensuring consistency across diverse platforms.

In our previous work~\cite{fkmm_nvidia_tc} we investigated the numerical features of NVIDIA tensor cores.
Using a combination of test vectors targeting specific floating-point behaviours and the generalised numerical feature testing (GNFT) framework from~\cite{khmi25}, we identified a wide range of numerical features. While most features were extracted using GNFT, some required ensemble-based random testing, revealing new features that require the refinement of the models.
This study covered multiple architectures, including Volta, Ampere, Ada Lovelace, Hopper,
and Blackwell. Based on these findings, we developed MATLAB-based models that achieve bit-level accuracy and allow both architecture-specific and user-defined configurations.

More recently,~\cite{xie25_mmasim} investigated numerical features of both NVIDIA and AMD GPUs. However, their study reports results for various architectures without providing feature-targeted test vectors, making it difficult to independently verify the reported numerical behaviours and assess the reproducibility of the results.

Our contributions are as follows.
\begin{enumerate}
\item We investigate matrix multipliers on AMD GPUs across CDNA 1 (MI100), CDNA 2 (MI210, MI250), and CDNA 3 (MI300A, MI300X) architectures. Where previous literature~\cite{xie25_mmasim} only specified the behaviour, we provide a detailed characterisation of their numerical features, together with test vectors and explanations of their effectiveness. This enables the testing methodology to be extended to future AMD architectures.  
Our findings also reveal differences from the characterisation reported in~\cite{xie25_mmasim}, which we discuss in detail in the paper. 
The repository accompanying~\cite{xie25_mmasim} on GitHub\footnote{\url{https://github.com/microsoft/MMA-Sim}}, however, is consistent with our findings.

\item We develop MATLAB-based software models for all three CDNA architectures with flexibility to alter various features for numerical experimentation. We expanded the MATLAB Tensor Core software, which previously consisted of more than 10 models of NVIDIA GPU tensor cores~\cite{fkmm_nvidia_tc}, with CDNA matrix cores.
\item We demonstrate the impact of these differences using  multi-word arithmetic for emulating high-precision GEMM via low precision~\cite{mami25} and broadband signal-processing algorithm~\cite{SMD}, where GEMMs constitute the dominant computational workload.
\end{enumerate}

\section{Notations and Definitions}
\label{sec_notation}

Table~\ref{table:fp-formats} shows characteristics of various floating-point formats available on the latest AMD architecture.
Hereafter we refer to floating-point formats with the following short-hand names: fp8 (either of E4M3 or E5M2), fp16 (binary16 of IEEE 754~\cite{ieee19}), bf16 (bfloat16), tf19 (commonly referred to as tensorfloat32), fp32 (binary32 of IEEE 754), and fp64 (binary64 of IEEE 754).
As a side note, AMD documents refer to E5M2 and E4M3 by fp8, and bf8, respectively.

\begin{table}[t]
  \begin{center}
    \caption{Floating-point formats which are available as input formats to matrix cores, in the GPU devices based on the AMD CDNA 1--4 architectures. CDNA 4 is included for completeness, but we do not have access to this architecture at the time of writing.}\label{table:fp-formats}
  \begin{tabular}{lrlrlrlc}
    \toprule
    Format & precision ($f$)  & min norm. pos. & max pos. & CDNA Arch. \\
    \midrule
    binary64~(double) & 53 & $2^{-1022}$ & $\sim 1.798 \times 10^{308}$ & 1/2/3/4  \\
    binary32~(single) & 24 & $2^{-126}$ & $\sim 3.403 \times 10^{38}$ & 1/2/3/4 \\
    tf19 (19-bit) & 11 & $2^{-126}$ & $\sim 3.401 \times 10^{38}$ & 3/4\\
    bfloat16 & 8 & $2^{-126}$ & $\sim 3.389 \times 10^{38}$ & 1/2/3/4\\
    binary16~(half) & 11 & $2^{-14}$ & $65504$ & 1/2/3/4\\
    fp8-E4M3 (FNUZ/OCP) & 4 & $2^{-7}/2^{-6}$ & $240/448$ & 3/4\\
    fp8-E5M2 (FNUZ/OCP) & 3 & $2^{-15}/2^{-14}$ & $57344/57344$ & 3/4\\
    \bottomrule
  \end{tabular}
  \end{center}
\end{table}

Take two matrices $A \in \mathbb{R}^{m\times k}$, $ B\in\mathbb{R}^{k\times n}$.
The AMD matrix cores accelerate a matrix fused-multiply-add (MFMA) operation
\begin{align*}
    D=AB+C \in \mathbb{R}^{m\times n}.
\end{align*}
In the scope of MFMA, $A$, $B$ are input matrices and $C$, $D$ are accumulator and output matrices, respectively.
For simplicity, if we denote with $d_{ij}$ the element at $i$th row and $j$th column of $D$,
 we can express $d_{ij}$ as the inner product between the $i$th row of $A$ and $j$th column of $B$ and an addition of a corresponding element in $C$ as
\begin{align*}
    d_{ij}=\sum_{\ell=1}^{k}a_{i\ell}b_{\ell j}+c_{ij}.
\end{align*}
To focus on the inner product as an underlying operation, rather than on particular elements of $D$, we will omit the subscripts:
\begin{align}
  \label{eq:inner}
    d=\sum_{\ell=1}^{k}a_{\ell}b_{\ell}+c=\sum_{\ell=1}^{k}p_{\ell}+c.
\end{align}

The operations in \eqref{eq:inner} can be implemented either as a sequential fused multiply--accumulate (SFMA) or using a hardware-supported multi-term floating-point addition strategy~\cite{mika24,fkmm_nvidia_tc,tenc09,aldi25}. Moreover, multiple implementations of multi-term floating-point addition exist~\cite{fkmm_nvidia_tc,xie25_mmasim}. Such implementations may employ additional alignment bits during significand alignment, which we denote by $\neab$.
For architectures that accumulate the $c$ term early, $\neab$ directly represents the number of additional alignment bits. In contrast, for architectures that accumulate $c$ after the product accumulation (late accumulation), this interpretation does not apply directly. In this case, $\neab$ refers only to the alignment of the product terms, whereas the alignment of $c$ with the accumulated product sum, whether normalized or denormalized, is described explicitly for each architecture.
Throughout this work, for any many-term dot-product accumulation, we define the number of product terms accumulated in a single multi-term addition, regardless of late or early $c$ accumulation, as the \emph{FMA size} ($\nfma$)~\cite{fkmm_nvidia_tc}. 
Throughout this paper, the term \textit{fractional bits} is used exclusively to denote the fractional-bit component of a fixed-point representation.
%
%
\section{Hardware-Guided Numerical Feature Detection and Model Refinement}
Numerical feature testing~\cite{khmi25,llfs24,fhmp21,hibr19} relies on a predefined numerical feature space and then uses targeted feature test vectors to determine whether particular features are present in a given matrix or tensor core architecture. While features may naturally exhibit interdependencies, targeted feature testing alone cannot be considered entirely reliable. 
Moreover, if a feature present in the hardware is not included in the feature space, it cannot be identified by targeted tests restricted to that space.

To address this limitation, we designed~\cite{fkmm_nvidia_tc} an iterative model refinement strategy that combines numerical feature testing with randomized testing against actual hardware. The approach constructs a model from the currently identified features and compares its outputs with those of the hardware over carefully selected randomized test vector sets. A mismatch indicates that the current feature space or model is incomplete, prompting further investigation and the introduction of additional features into the refinement process. This iterative procedure continues until the model agrees with the hardware for the test vectors considered.

For the proposed modelling framework, we construct an initial comprehensive feature space by incorporating all features considered in~\cite{khmi25}, together with additional features identified in NVIDIA GPU tensor core architectures through randomized testing in~\cite{fkmm_nvidia_tc}. We also include all features reported for AMD GPUs in~\cite{xie25_mmasim}. These features include, among others, product and accumulation precision, rounding mode, and the number of alignment bits used when aligning product significands.
Because the input space of matrix core operations is prohibitively large to explore exhaustively, randomized testing requires an intelligent sampling strategy to efficiently identify discrepancies between the hardware and the model. 
Such discrepancies provide evidence of potentially missing numerical features and guide subsequent refinement of both the feature space and the model. 
In Section~\ref{sec:EnsTest}, we discuss how such a finite set of test vectors is constructed.
\section{Results}
We have analyzed compute DNA (CDNA) 1, 2 and 3 architectures in this paper by running numerical feature detection similar to~\cite{khmi25}.
We have used MI100, MI210, MI250, MI300, and MI300X AMD data centre GPUs.
In all CDNA architectures, matrix cores can be invoked via specific intrinsic MFMA instructions: \texttt{mfma\_In\_shape\_Out} where \texttt{In} can take on fp32, fp16, and bf16 input formats while \texttt{Out} is strictly in fp32, and matrix \texttt{shape} is in format \texttt{mxnxk} which can take on values that vary with input formats.\footnote{Examples: \texttt{\_\_\_builtin\_amdgcn\_mfma\_f32\_16x16x4f32}, \texttt{\_\_\_builtin\_amdgcn\_mfma\_f32\_16x16x16f16}}

\subsection{CDNA 1 Architecture}
We address each input format case separately.

\subsubsection{CDNA1 fp32 input} The supported matrix sizes for the \texttt{shape} argument are \texttt{32x32x1} \texttt{16x16x1}, \texttt{4x4x1}, \texttt{32x32x2} and \texttt{16x16x4}.
    We adopted the identical numerical feature determining strategy as that of the previous work~\cite{fhmp21,khmi25,llfs24} and we detected that fp32 input matrix cores are SFMAs (see Section~\ref{sec_notation}).
    Initially, the rounding mode is determined from $p_1$, $c$, and the resulting $d$, as defined in \ref{eq:inner}. 
    For $p_1=\pm1$ and $c=\pm\{2^{-23}+2^{-24},\,2^{-24}\}$, with all inputs having the same sign, $d=\pm\{1+2^{-22},\,1\}$, indicating round-to-nearest, ties-to-even (RNE). 
To determine whether the operation is an SFMA-like, 
we first permuted $1$, $2^{-23}$, and $2^{-24}$ among $p_1$, $p_2$, and $c$. We obtained $d=1+2^{-23}$ when $p_2=2^{-23}$; otherwise, we obtained $d=\pm(1+2^{-22})$. These results suggest that $c$ and $p_1$ are added first, and their sum is then added to $p_2$.
We then permuted $1$, $2^{-23}+2^{-24}$, and $2^{-23}+2^{-24}$ among $p_1$, $p_2$, and $p_3$. For all permutations except the case $p_1=p_2=2^{-23}+2^{-24}$ and $p_3=1$, we obtained $d=1+2^{-21}$. For this exceptional case, we obtained $d=1+2^{-22}+2^{-23}$. 
These results demonstrate that the fp32 input matrix core internally implements an SFMA for all supported matrix sizes, i.e., $(\dots(((c+p_1)+p_2)+p_3)+\dots)$.
    Although, in an SFMA, it can be difficult to determine how products are computed and kept, we still perform a series of following tests.
    \begin{enumerate}
    \item By setting $p_1=\pm(2^{-149}+2^{-150})$ via $a_1=\pm(2^{-126}+2^{-127})$, $b_1=2^{-23}$, and $p_i=c=0$ for all $i\neq 1$, we obtain $d=\pm 2^{-148}$. This result indicates that either the products are rounded to FP32 using RNE, or the RNE operation is applied only at the end of the SFMA operation.

    \item Next, we set $p_1=2^{-149}$ and $p_2=2^{-149}+2^{-150}$, and obtain $d=2^{-148}$. This result confirms that the previous observation was caused by the final rounding step after the SFMA operation. Otherwise, if the products were rounded before the addition, we would have obtained $d=2^{-148}+2^{-149}$. Therefore, this indicates that the intermediate products are retained in full precision before the addition stage.

    \item For further confirmation, we set $p_1=2^{-149}$ and $p_2=2^{-148}+2^{-150}$ to prevent any possible upward rounding of the products. We obtain $d=2^{-147}$, which further confirms that the rounding observed in the first case was due to the final rounding step. Hence, the products are preserved in full precision throughout the SFMA operation before the final rounding.
\end{enumerate}
    Both positive and negative Inf values are supported in the output, i.e., with $a_1=a_2=\pm \mathrm{max}_{\mathrm{fp32}}$, which is $(2-2^{-23})2^{127}$, and $b_1=b_2=1$, the output is $\pm$Inf.
    However, when positive and negative infinities are added, the resulting NaN carries a negative sign i.e., $a_1=-a_2=$ Inf, $b_1=b_2=1$, the output is $-$NaN. 
    Subnormal values are supported in both input and output.
    For example, when $a_1=\pm2^{-127}$ and $b_1=1$, the output is $d=\pm2^{-127}$. Similarly, when $a_1=\pm b_1=2^{-65}$, the output is $d=\pm2^{-130}$, which is a subnormal value rather than zero. 
\subsubsection{CDNA1 fp16 input}
    In fp16 input format, the supported matrix sizes are \texttt{32x32x4}, \texttt{16x16x4}, \texttt{4x4x4}, \texttt{32x32x8} and \texttt{16x16x16}.
    We first determined that subnormals are supported both in input, because we obtained $d=\pm 2^{-48}$ for $a_1=2^{-24}$, $b_1=\pm 2^{-24}$, and also in the output by producing $d=c$ for $c=\pm 2^{e}$ for $e\le -126$.
    Then we have determined the final rounding mode to be RNE using the same test vector as used in the case of fp32 input format.
    The SFMA test produced $d=1+2^{-22}+2^{-23}$ for all permutations which reflects the absence of an SFMA structure.
    Moreover, we permuted $1$, $2^{-23}+2^{-24}$ between $c$ and $p_1$, while $p_j=2^{-23}+2^{-24}$ for $j>1$ while $p_\ell=0~\forall \ell\neq j\And \ell\neq 1$.
    When $j$ was varied from $2$ onward, we obtained $d=1+2^{-22}+2^{-23}$ for $j\le 4$ and $d=1+2^{-21}$ for $j>4$, an SFMA-like
    behaviour, which revealed that the $\nfma=4$ for all supported matrix sizes. This test differs from the $\nfma$ detection algorithm in \cite{khmi25}, as the final rounding mode  in CDNA 1 is detected to be RNE instead of RTZ.
    The permutation aspect shows that $c$ is added along with the products, via simultaneous alignment of significands, revealing an NVIDIA tensor-cores like behaviour~\cite{fkmm_nvidia_tc}.

    In multi-term floating point addition~\cite{mika24} where significands are aligned w.r.t the largest exponent, we expect to see finite number of alignment bits beyond the output precision fractional bits.
    To determine this feature, we performed the following tests:
    \begin{enumerate}
    \item we permuted $1$, $2^{-24}$, $2^{-25-\neab}$ between $c$, $p_1$, $p_2$, a similar approach as that of~\cite{khmi25,hibr19}, where $\neab$ is varied from $0,\dots,24$ because the minimum value of product exponent is $-48$ in fp16 input format,
    we found $d=1+2^{-23}$ for every value of $\neab$.
      This showed that either there are multiple sticky bits~\cite{tenc09}, or a correctly rounded implementation that adds all products in full precision (a Kulisch accumulator~\cite{kulisch2008computer}), is present.
        \item by permuting $1$, $2^{-23}+2^{-24}$, $-2^{-25-\neab}$, $d$ remains $1+2^{-23}$
        for all possible values of $\neab\ge 0$.
        The output remains unchanged even when $c=-2^{-25-\neab}$,
        which is in fp32 format, and therefore, $\neab$ is varied
from $0,\dots,(149-25)$ due to the subnormal support.
      \item we permuted $\pm1$, $\pm2^{-23}$, $\pm\sum_{\ell=-25}^{L}2^{-\ell}$, $\pm2^{-L}$ where $L$ is varied from $-26$ to $-48$,
      for product computed with two fp16 format elements.
        We obtained $d=\pm(1+2^{-22})$ for all cases. Note all elements are either positive or negative.
    \end{enumerate}
The results of these tests reveal full precision addition or correctly rounded accumulation, and therefore, theoretically we can say that $\neab=\infty$. 
These results are contrary to
that reported by~\cite{llfs24} that MI100 matrix cores have $3$ extra alignment bits.
Unlike the NVIDIA tensor cores, the denormalised-product feature is not relevant here because the accumulation is performed accurately. Consequently, this feature may not be directly testable, but it is also unnecessary, since $\neab = \infty$.
Both positive and negative infinities are detected in the input i.e. $a_1=\pm$Inf, $b_1=1\Rightarrow d=\pm $Inf, and in case of both positive and negative infinities, the output is NaN with a negative sign i.e. $a_1=-a_2=$ Inf, $b_1=b_2=1\Rightarrow d=-$NAN.
Within the fp16 input format, the product exponent can reach a minimum value of $-28$, and with subnormal inputs,
it can be as low as $-48$. Therefore, this case cannot be verified if the product is rounded to fp32. However, for the bf16 input format, this verification has been performed, as discussed further below.

\subsubsection{CDNA1 bf16 inputs} The supported matrix sizes are \texttt{32x32x2}, \texttt{16x16x2}, \texttt{4x4x2}, \texttt{32x32x4},~and \texttt{16x16x8}.
In all supported sizes, we were able to determine that $\nfma=2$ performing identical test as carried out for fp16 case. 
Similarly, subnormals both in input and output are also supported with final rounding as RNE.
The product of two bf16 numbers has a significantly wider dynamic range than a single bf16 number, as their exponents add during multiplication, resulting in an approximate range from $2^{-266}$ to $2^{254}$, therefore, we determine if products are kept in full precision instead of rounding to fp32 before addition.
To determine this, we set $p_1=\pm(2^{-149}+2^{-150})$, $p_2=\pm2^{-149}$, and then $p_1=\pm(2^{-148}+2^{-150})$, $p_2=\pm2^{-149}$, with both products residing within one block FMA, we obtained $d=\pm2^{-148}$, and $d=\pm 2^{-147}$, respectively, for both cases.
This shows that products are kept in full precision as if it were not the case, and the products were rounded to fp32 format via RNE before addition, we would have obtained $d=\pm(2^{-148}+2^{-149})$ for both cases.
For other types of rounding of the product to fp32 format, the output would still be different.
In addition, we set $c=\pm 1$, $p_1=\pm(2^{-23}+2^{-24})$, and $p_2=\mp 2^{-25-\neab}$, where $c$ and $p_1$ have opposite signs to $p_2$,
and $\neab=0,\dots,(266-25)$,
and we obtained $d=\pm(1+2^{-23})$ which further confirms that products are computed and stored as the sum of exponents multiplied by the product of significands in full precision. 

Infinities are detected at the bf16 input stage, even when the corresponding product would not overflow in fp32 precision. 
For example, $a_1 = \pm 2^{128}$, which represent infinity in bf16, and $b_1 = 2^{-20}$
yields $p_1 = \pm 2^{108}$,
which is representable in fp32.
Nevertheless, the computed result is $d = \pm \infty$.
However, if the inputs are not infinities, the intermediate products are allowed to grow beyond \(2^{128}\) without being mapped to infinities. For instance, we obtained $d = 0$
for
$p_1 = 2^{129}$, $p_2 = -2^{129}$,
generated via
$
a_1 = 2^{65}$, $b_1 = 2^{64}$,
and
$a_2 = 2^{65}$, $b_2 = -2^{64}$.
This suggests that infinity detection occurs at the input level, whereas overflow during intermediate product accumulation is handled at the very end when sum is converted to fp32.
On the opposite end of the dynamic range, where the product falls below the smallest normal fp32-representable magnitude, we observe from the output that $d = \pm 2^{-148}$
for
$c = \pm 2^{-149}$,
$p_1 = \pm \sum_{\ell=151}^{156} 2^{-\ell}$,
$p_2 = \pm 2^{-156}$.
This indicates that products smaller than $2^{-149}$ still participate in the computation and are not flushed to zero. 
Because CDNA 1 matrix cores manifest full-precision accumulator within the dot products, many of the features later tested on CDNA 2 and CDNA3, as well as those tested on NVIDIA tensor cores \cite{fkmm_nvidia_tc}, do not apply.
With both $+$Inf and $-$Inf in inputs, the output is $-$NaN, consistent with fp32 and fp16 input formats.

\subsubsection{CDNA1 summary}
\begin{figure}
    \centering
    \includegraphics{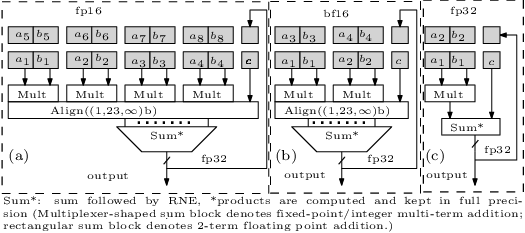}
    \caption{CDNA 1 matrix core model diagram. (a) fp16 input format with an fma size of $4$, (b) for bf16 input format with an fma size of $2$, and (c) fp32 input format which is an SFMA. All products are in full precision, the sum is of full accuracy, and final result is rounded to fp32 via RNE.}
    \label{fig:CNDA1}
  \end{figure}

The entire matrix core block diagram for CDNA 1 arch is illustrated in Fig.~\ref{fig:CNDA1}, where multiplication blocks represent full precision products.

  \subsection{CDNA 2 Architecture}

CDNA 2 matrix cores have been investigated previously by~\cite{xie25_mmasim}, but without test vectors provided.
Here we are providing exact test vectors and confirm the details of the matrix cores.
We have observed behaviour in following input formats:
  The MFMA intrinsic instructions support the same matrix sizes as those in the CDNA1 architecture for fp32 inputs. 
  All other features are identical to the CDNA1 matrix cores with fp32 input format.
    
\subsubsection{CDNA2 fp16 input}

    The supported matrix sizes for fp16 are the same as those in the CDNA1 architecture.
    Subnormals are not supported neither in input nor in the output. This means that if products are zero, and $c$ is less than $2^{-126}$, output is zero. Similarly, for $c=0,~a_1=2^{-24}$ and $b_1=1$,
    we obtained $d=0$. 
    The final rounding mode is detected to be RNE from $d=\pm\{1$, $1+2^{-22}\}$ for $c=\pm\{ 2^{-24}$, $2^{-23}+2^{-24}\}$, $p_1=\pm 1$.
    The addition of multi-term floating point mechanism is detected via the following tests:
    \begin{enumerate}
    \item permuting the values $1,~2^{-23}+2^{-24}$, $2^{-23}+2^{-24}$ between $c$, $p_1,$ and $p_2$ produced different values of $d$ indicating absence of global alignment between these three mentioned terms.
        \item with $c=1$, $p_1=p_j=2^{-23}+2^{-24},~p_\ell=0$, where $\ell\neq j\And \ell\neq 1$, for $j<5$, we obtained $d=1+2^{-22}+2^{-23}$ while for $j>4$, it produced $d=1+2^{-21}$ indicating $\mathrm{fl}(c+\mathrm{fl}(p_1+p_j))$ for $j<5$. 
        \item with $p_1=1$, $p_4=p_3=2^{-23}+2^{-24}$,
        all remaining factors set to zero, we obtained $d=1+2^{-22}+2^{-23}$. 
        With $p_2=1,p_4=p_3=2^{-23}+2^{-24}$ produced same results indicating $\fl(p_j+\fl(p_3+p_4))$ for $j<3$.
        \item lastly, $p_1=p_3=1$, $p_2=p_4=2^{-23}+2^{-24}$ produced $d=2+2^{-21}$ confirming $\fl(p_1+p_2)+\fl(p_3+p_4)$ while additionally setting $c=2^{-23}+2^{-25}$ resulted in $d=2+2^{-21}+2^{-22}$ confirming a pair-wise sum in groups
        which follows computations: $\fl(c+\fl(\fl(p_1+p_2)+\fl(p_3+p_4)))$.
    \end{enumerate}
 This shows that pair wise sum is computed in sizes of $4$ which confirms the results reported by~\cite{xie25_mmasim}. 
Moreover, at each RNE operation, it is unclear whether a pairwise sum that becomes an fp32 subnormal value is preserved or flushed to zero. This behavior cannot be tested using fp16 inputs because such values cannot be represented, but it is confirmed later using bf16 inputs.
 The grouping size is $4$ for all supported matrix sizes.
 Similarly, infinities are detected at the input level, and the output is NaN with a negative sign when both positive and negative infinities are present in the input, i.e. $a_1=a_2=1,b_1=-b_2=$ Inf, $\Rightarrow d=-$NaN.


\subsubsection{CDNA2 bf16 input}

\begin{figure}
    \centering
    \includegraphics{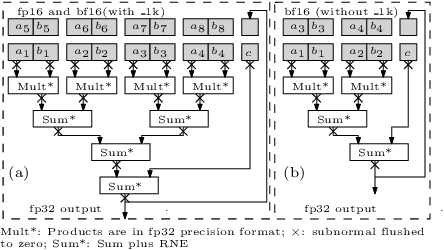}
    \caption{CNDA 2 matrix core model diagram. (a) shows fp16 and bf16 (with subscript \texttt{\_1k}) input format with a group size of $4$, and (b) for bf16 input format, without \texttt{\_1k} with a group size of $2$. fp32 input format matrix cores are identical to the ones in the CDNA 1 architecture, and are not shown.}
    \label{fig:CDNA2}
  \end{figure}

Unlike CDNA1, and also the fp16 input format in CDNA2, the bf16 format introduces an additional subscript \texttt{\_1k} in the MFMA intrinsic instruction for certain input matrix sizes. The matrix sizes that include this additional subscript are \texttt{32x32x4}, \texttt{16x16x4}, \texttt{4x4x4}, \texttt{32x32x8}, and \texttt{16x16x16}. The matrix sizes without this subscript are the same as those supported in the CDNA1 architecture.
Similar to the fp16 case, the final rounding mode is RNE. 
The group-wise pair-sum test vectors listed for the fp16 case indicate a group size of $4$ for the intrinsic instructions with the subscript \texttt{\_1k} (see Figure~\ref{fig:CDNA2}(a)), whereas for the remaining instructions the group size is $2$ as depicted in Figure~\ref{fig:CDNA2}(b).

Regarding the product precision, which was not possible in the fp16 input case, we evaluated the case $p_1=2^{128}$ and $p_2=-p_1$. The resulting output was $d=-\mathrm{NaN}$, indicating that the products are converted to fp32 before accumulation rather than being accumulated in full precision as in the CDNA 1 architecture.  
Subnormals are flushed after every $\fl\{\cdot\}$ operation i.e. after each normalisation followed by RNE. 
For instance, $a_1=b_2=2^{-64}$, $a_2=b_1=2^{-63}$, which makes the product subnormal i.e., $p_1=p_2=2^{-127}$, the output remains $d=0$.

Similar to the fp16 case, both positive and negative Infs are detected in inputs $a$, $b$, and $c$; for $a_1=1,~b_1=\pm$Inf, we obtain $d=\pm$Inf, whereas for $a_1=a_2=1,~b_1=-b_2=\pm$Inf, we obtain $d=-$NaN. Figure~\ref{fig:CDNA2}[b] illustrates the matrix core structure corresponding to the bf16 input format.

\subsection{CDNA 3 Architecture}
\label{sec_cdna3_arch}
For the CDNA 3, we used MI300, MI300A and MI300X AMD GPUs for experiments.
The features of matrix multipliers in the CDNA 3 are also reported by~\cite{xie25_mmasim}; however, no detailed explanation of how they were determined or test vectors are provided.
Unlike CDNA 1 and 2, CDNA 3 provides support for fp8 inputs in the matrix cores.
    
    In terms of fp32 input, the matrix sizes supported in CDNA 1 and CDNA 2 are also supported in the CDNA 3  behave as SFMAs. 
\subsubsection{CDNA3 fp16 input}
\label{subsec_cdan3_fp16}
The matrix sizes supported are the same as those in CDNA 1 and 2.
Subnormal numbers are supported in both input and output, and the final rounding mode is RNE. 
On this architecture, we first tested for the existence of extra alignment bits by permuting $\pm 1,~\pm 2^{-24}$, and $\pm 2^{-24}$ between $c$, $p_1,$ and $p_2$.
The result $d = \pm(1 + 2^{-23})$ suggests presence of an extra alignment bit. 
This also at least indicates that $c$ is not added after normalization of the product sum. 
Otherwise, under the assumption of RNE rounding, the output would have been $\pm1$ for $p_1=\pm1$ and $p_2=c=\pm2^{-24}$.
However, the stage at which $c$ is added to the product sum remains to be determined and is investigated through the additional tests presented further below.

Next, given $\neab\geq1$, we perform the FMA size detection test with $c=1$, $p_1=2^{-24}$,  $p_j=2^{-24}$ and $p_{i\neq 1 \And i\neq j}=0$. We obtain $d=1+2^{-23}$ for $j\leq8$, indicating an $\nfma=8$, which confirms the results reported in~\cite{xie25_mmasim}. 
To rule out the possibility that odd- and even-indexed products are added separately i.e. inter-leaved pattern as detected in~\cite{fkmm_nvidia_tc},
we permuted the values $\pm 1$, $0$, and $\pm(2^{-24} + 2^{-25})$ across $p_1$, $p_2$,
and $p_3$. 
The resulting outputs were invariant, consistently yielding $d = \pm 1$ which implies no interleaving. 
To confirm whether $\neab=1$ or $\neab>1$, we used the test vector
$
p_1 = \pm 1$, $
p_2 = \pm (2^{-24}+2^{-24-j})$.
For $j>0$, we obtained $d = \pm 1$ indicating that bits beyond the 24th fractional bit are truncated during alignment. 
Since the final rounding mode is RNE, if bits beyond $24$th bit were not truncated, the output would not be $\pm 1$. For instance, if bits beyond $25$th bit were truncated, $d$ would have been $\pm(1+2^{-23})$. Moreover, if it was a round up or down, or RNE at the alignment, $d$ cannot be equal to one for both positive or negative signs.
With a single extra alignment bit, we tested whether a feature similar to NVIDIA tensor cores exists, where products remain denormalized. 
Using $p_1=2.25$, $p_2=2^{-23}$, and $p_3=p_4=2^{-24}$, we observed \(d=2.25+2^{-22}\) for \(a_1=b_1=1.5\). 
In contrast, when $a_1=2.25$ and $b_1=1$, we obtained \(d=2.25\).
This suggests that products are computed as the product of significands multiplied by $2^{(\text{sum of exponents})}$ and are kept denormalised while being accumulated. 
To check whether $c$ is aligned and added together with the products, we set $c = 1$, $p_1 = p_2 = 2^{-25}$, and $p_3 = 2^{-24}$. 
We obtained $d = 1 + 2^{-23}$, indicating that $c$ is added after product accumulation; otherwise, we would have obtained $d = c$ because $\neab=1$ and final rounding mode is RNE.
To further analyze the alignment of $c$, we used
$p_1 = \pm 1$, $c = \pm (2^{-24}+2^{-24-{j}})$
where $j > 0$.
We observed:
\begin{enumerate}   
\item for positive values, $d = 1$ for $j> 0$ indicates the truncation of bits beyond the 24th fractional bit when the significand of $c$, denoted with $s_c$, is shifted w.r.t the products' significands sum, denoted as $S_{p_i,\mathrm{sum}}$. Additionally, using $p_1 = 1$, $p_2 = 2^{-24}$, and $c = 2^{-24}$, we obtain $d = 1 + 2^{-23}$, confirming that the product sum is not rounded to output precision before $c$ is added.
\item on the negative axis, we obtained $d = -(1 + 2^{-23})$ for all $j>0$, indicating that the shifted $s_c$ is rounded to $24$ fractional bits using RD. 
\end{enumerate}

These results are consistent with the observations reported in~\cite{xie25_mmasim}, although explicit test vectors are not provided in that work.
The fractional bit width when products sum, which is not normalized before adding to $c$, is shifted against $c$, we performed following tests:
\begin{enumerate}
 \item $c=\pm1$, $p_1=\pm(2^{-24}+2^{-24-j})$, with $j>0$. We observed $d=1+2^{-23}$ for $j<8$ and $d=1$ for $j\ge8$ in the positive case, and $d=-1-2^{-23}$ for all $j>0$ in the negative case. 
 In isolation, these results appear to suggest that the shifted $\psumsig$ is rounded down to $31$ fractional bits.
   \item Due to mismatches in randomized testing discussed in Section~\ref{subsec_testing}, we performed an addition test with $c=1$, $p_1+\dots+p_{\nfma}=-(2^{-24}+\sum_{\ell=26}^{30+j}2^{-\ell})$, where $j$ is varied from $0$ to $10$. 
   We obtained $d=1-2^{-24}$ for $j\leq2$ and $d=1-2^{-23}$ for $j>2$. 
   This test reflects that the shifted $\psumsig$ is rounded-down to $32$ fractional bits instead to $31$ fractional bits.
   \item 
   To further validate this conclusion and resolve the apparent contradiction with the previous test, we set $c=1$, and chose
$p_1+\dots+p_{\nfma}=-(2^{-24}+\sum_{\ell=26}^{31}2^{-\ell})+2^{-33}$
where the $32$nd bit is intentionally set to $0$ to prevent the carry generated by RD from propagating to the $25$th fractional bit.   
We obtained $d=1-2^{-24}$, confirming that alignment-stage RD is applied at the $32$nd fractional bit of the shifted $\psumsig$. 
Assuming this is the case, the sum of the aligned $s_c$ and $\psumsig$, denoted by $S_{\mathrm{acc}}$, is $S_{\mathrm{acc}}=1-2^{-24}-\sum_{\ell=26}^{32}2^{-\ell}$, which results in $d=1-2^{-24}$ after RNE. 
In contrast, if RD were applied at the $31$st fractional bit, then $S_{\mathrm{acc}}=1-2^{-24}-\sum_{\ell=26}^{31}2^{-\ell}-2^{-31}=1-2^{-24}-2^{-25}$ which would produce $d=1-2^{-23}$ after RNE, contrary to the observed result.
 
\end{enumerate}
The observation that the shifted $\psumsig$ is rounded down to 32 fractional bits contradicts the 31 fractional bits reported in~\cite{xie25_mmasim}, yet it surprisingly matches the behaviour implemented in the associated repository.
The observation from earlier test that $d=1+2^{-23}$ for $j<8$ and $d=1$ for $j\ge8$ cannot be attributed to a $31$-bit alignment precision. 
With alignment-stage RD confirmed to retain $32$ fractional bits, the apparent $31$-bit behaviour instead arises from a subsequent RD operation applied to the normalised accumulated significand $\Sacc$ before the final RNE step. 
This observation was made due to mismatch in the randomized testing.
 
To characterize this second RD operation, we investigate whether $\Sacc$ is truncated or rounded to a $32$-bit representation before the final RNE stage by executing the following tests:
\begin{enumerate}
  \item $c=2-2^{-22}$, $p_1=2^{-22}$, and $p_2=2^{-23}+2^{-j}$, where $j$ is varied from $24$ to $32$. 
  This produces $d=2+2^{-22}$ for $j\leq30$ and $d=2$ for $j>30$, seemingly suggesting that the RD at the alignment stage operates at the $30$th fractional bit. 
  However, with the RD at the $32$nd bit confirmed, this behavior strongly indicates the presence of an additional rounding operation on $\Sacc$ before the final RNE. 
  For $j=31$ and $32$, applying RD at the $32$nd bit yields $\psumsig=2-2^{-23}+2^{-j}$. If $\Sacc$ is truncated to a total of $32$ bits at this stage, the resulting values are consistent with the observed results.
    \item Additionally, we set $c=-2+2^{-21}$, $p_1=-2^{-21}$, $p_2=-2^{-22}$, and $p_3=-\sum_{\ell=24}^{j}2^{-\ell}$, where $j$ is varied from $28$ to $32$. 
    We obtain $d=-2-2^{-22}$ for $j=28,\ldots,30$ and $d=-2-2^{-21}$ for $j>30$.
It further confirms that the alignment stage RD occurs at the $32$nd fractional bit after $\psumsig$ is shifted. 
In addition, it also suggests that $\Sacc$ is rounded down, rather than simply truncated to 32 bits.
For example, consider the case $j=31$. 
Assuming the shifted $\psumsig$ is rounded down to $32$ fractional bits, the $\Sacc$ would be $-2-2^{-22}-\sum_{\ell=24}^{31}2^{-\ell}$, which has 2 integer and 32 fractional bits. 
Truncation of $\Sacc$ to $32$ bits would result $-2-2^{-22}-\sum_{\ell=24}^{30}2^{-\ell}$, and RNE, $d=-2-2^{-22}$. 
However, the observed result is $d=-2-2^{-21}$ which indicates that the final sum of significands is rounded down to 32 bits rather than simply truncated. 
\end{enumerate}

In both tests above, \(\Sacc\) has \(2\) integer bits and \(32\) fractional bits i.e.,
\begin{align*}
\Sacc = 2 + 2^{-23} + 2^{-31}=10&.00000000000000000000001000000010~\text{Test\#1 for $j=31$} \\
\Sacc =-2-2^{-22} -\sum_{\ell=24}^{31}2^{-\ell}=-10&.00000000000000000000001011111110~\text{Test\# 2 for }j=31
\end{align*}
However, the post-RD operation on $\Sacc$ appears to retain only $30$ fractional bits. 
In the positive case, the last 2 bits
are discarded, whereas in the negative case the result is rounded down to $-2-2^{-22}-2^{-23}$.
Now consider an earlier test with \(c=1\), \(p_1=2^{-24}\), and \(p_2=2^{-j}\). 
This test produced \(d=1+2^{-23}\) for \(j=31\) and \(d=1\) for \(j=32\). 
In this case,
\begin{align*}
\Sacc &= 1 + 2^{-24} + 2^{-31}=1.00000000000000000000000100000010 \\
\Sacc &=1 + 2^{-24} + 2^{-32}=1.00000000000000000000000100000001
\end{align*}
which clearly requires \(1\) integer bit and $31$ and $32$ fractional bits, respectively. 
The RD operation appears to preserve \(31\) fractional bits. 
This behaviour suggests that the RD operation applied to \(\Sacc\) may be accompanied by a small normalization step.
To confirm this, we set $c=1$ and $p_1+\dots+p_\nfma=-(2^{-24}+\sum_{\ell=26}^{32}2^{-\ell})$ 
which yields in 
\begin{align*}
\Sacc=1-2^{-24}-\sum_{\ell=26}^{32}2^{-\ell}=0.\underbrace{111111111111111111111110}_{\text{24 bits}}10000001.    
\end{align*} 
If the RD operation on $\Sacc$ were performed blindly, i.e., without a normalisation, then the last $1$ bit would have to be discarded, as in the earlier case where $31$ fractional bits remained. Consequently, we would obtain $d=1-2^{-23}$.
However, since the actual value of $d$ is $1-2^{-24}$ which shows the RNE operation has not rounded up the output, confirms that the RD operation is applied after the normalisation of $\Sacc$, and therefore not discarding the last bit. 
This constitutes the main normalisation step. 
In addition, two smaller normalisation steps are performed: one before and one after the final RNE operation.
We cannot test this intermediate normalisation before RD operation of $\Sacc$ for a subnormal $c$ in this format, as the minimum possible product exponent is only $-28\gg -126$.

Lastly, since the products are accumulated first and $c$ is accumulated afterward, we consider a cancellation test in which the $\psumsig$ becomes zero, but the maximum product exponent exceeds that of $c$ by more than 24, i.e.,  $c=2^{10}$ and $p_1=-p_2=2^{10+25}$. 
It produces $d=0$, indicating that $c$ is aligned w.r.t the maximum product exponent, even if $\psumsig$.
    
\subsubsection{CDNA3 bf16 input}
    All features are the same as that in fp16 input format.
    The product overflow feature could not have been validated in fp16 due to narrow range.
    However, in bf16, even though products are computed as product of significands times sum of the exponents raised to the base $2$, we obtained $\pm$Inf for $|p_1|\ge 2^{128}$. 
    Moreover, since the products are kept denormalised, for which we suspected to obtain $d=0$ for $a_1=a_2=1.5\times2^{64}$, $b_1=-b_2=1.5\times 2^{63}$, 
    but it produced $d=-$NaN which shows that even though the denormalised product's, $2.25\times 2^{127}$, exponent is below $128$, the overflow is detected based on the magnitude instead of only on the exponent. 
    However, an intermediate sum is allowed to exceed $2^{128}$ within one block FMA operation. 
    We further investigated whether a feature analogous to the exponent limitation observed in NVIDIA GPUs~\cite{fkmm_nvidia_tc} is present. 
    To this end, we conducted a series of tests by setting $c = 0$, $p_1 = \pm 2^{-150}$, and $p_2 = \pm 2^{-152-i}$, where $i$ is varied from $0$ to $23$, corresponding to the $24$ fractional bits. The following observations are made:
    \begin{enumerate}
        \item on the positive axis, it produced $d=2^{-149}$ for $i<6$, otherwise $d=0$. 
        This suggests that the minimum exponent used for aligning the significands, as reported in NVIDIA tensor core~\cite{fkmm_nvidia_tc}
        is $-132$. 
        \item on the negative axis, we obtained $d = -2^{-149}$ for all values of $i\ge 0$. This indicates that the product sum $p_1+\dots+p_{\nfma} = -2^{-150} - 2^{-152-i}$ is not directly rounded to the output precision as otherwise we would have obtained identical results to the above case but in negative; instead, it is first aligned w.r.t the maximum exponent of $-126$, even though $c = 0$. 
        This alignment yields $d = -2^{-149}$ for $i < 6$. For $i > 5$, terms such as $-2^{-158}$ or smaller fall beyond the $31$st fractional bit after alignment, and the product sum gets rounded down, which consistently results in $d = -2^{-149}$. 
        \item 
                The exponent limitation is not observed for zero-valued products. Specifically, when $p = 1 \times 0 = 0$, the behaviour is only restricted to $c = 0$.
    \end{enumerate}
   From these observations, we can confirm $c=0$ is considered as subnormal number as its exponent is considered to be $-126$.  
   Unlike NVIDIA GPU tensor cores, where $c=0$ is considered as a special case~\cite{fkmm_nvidia_tc}. 
   This feature of exponent of $c$, when it is zero, being -126, is not discussed by~\cite{xie25_mmasim}.
   However, this behavior is correctly implemented in the associated GitHub repository.

The normalisation step prior to RD application to $\Sacc$ when $c$ is a subnormal value could not be tested in fp16 input case. 
   Therefore, for this input format, we set $c=2^{-127}-2^{-140}$ while $p_1+\dots+p_{\nfma}=2^{-140}+2^{-150}+2^{-154-j}$ where $j$ is varied from $0$ to $10$.
For $j\leq3$, we obtain $d=2^{-127}+2^{-149}$, while for $j>3$, $d=2^{-127}$. 
Since the maximum alignment exponent is $-126$, $\Sacc$ has a zero integer bit; nevertheless, 31 fractional bits are retained, as $-126-(-154-j\le 3)=31$. 
Thus, the intermediate normalisation before RD operations is also subnormal aware.
   
    \subsubsection{XF32 (tf19)}
    In CDNA~3 additional MFMA instructions are available:
\texttt{f32\_16x16x8\_xf32} and \texttt{f32\_16x16x4\_xf32}. These instructions 
accept \texttt{float} inputs but internally reduce the mantissa to $10$ bits. 
This behaviour was verified experimentally by setting 
$a_1 = 1 + 2^{-10} + 2^{-11}$ and $b_1 = \pm 1$, which produced 
$d = \pm(1 + 2^{-10})$, reflecting truncation, instead of $\pm(1 + 2^{-9})$ which would reflect an application of RNE.
The $\nfma=4$ in this input format, which matches with that of Ampere and Ada family GPUs tensor cores while that for Hopper and Blackwell, it is 8. 
We also confirm that when $c=0$ the exponent of $c$ is equal to $-126$ in this input format as well.

\subsubsection{CDNA3 binary8 input}

In the binary8 input format, input matrices can be in any combination of E5M2 (fp8) and E4M3 (bf8), with matrix sizes \texttt{32x32x16} and \texttt{16x16x32}. 
Subnormals are supported at both input and output, and the final rounding mode is RNE. 
The detected FMA size is $16$.

Permuting $\pm1$, $0$, and $\pm(2^{-24}+2^{-25})$ among $p_1$, $p_2$,
and $p_3$ yields $d=\pm1$ when $p_2=0$. For other permutations, $d=1$ for positive values and $d=-(1+2^{-23})$ for negative values. 
This confirms the interleaving feature reported in~\cite{xie25_mmasim}. 
These results indicate that even- and odd-indexed products are accumulated separately, with truncation applied beyond $24$ fractional bits (i.e., $\neab=1$). 
The resulting partial sums are then accumulated without intermediate normalization, and bits beyond $24$ fractional bits are rounded down.

To analyse the addition of $c$, we permute the values $\pm1$ and $\pm\big(2^{-24}+2^{-24-i}\big)$ between $c$ and $p_1$, for $i=0,\dots,10$. Similar behaviour to the fp16 and bf16 cases is observed: The shifted significand is rounded down to 24 fractional bits when $c$ is shifted, and to 32 fractional bits when the accumulated odd-indexed product sum is shifted.
Repeating this experiment with permutations between $c$ and $p_2$ yields identical behaviour.
To further investigate the treatment of $c$ when it is shifted entirely beyond the retained precision of the product sum, we fix $p_1=\pm1$ and $p_2=2^{-24}$, while varying $c=\pm2^{-24-i}$. 
For $i=1$, the output is $d=1$ for positive inputs and $d=-(1+2^{-23})$ for negative inputs, consistent with the behaviour observed previously. 
However, for $i>1$, the result becomes $d=\pm1$ in both cases. This indicates that once the entire significand of $c$ is shifted beyond the $25$th fractional bit of the accumulated product sum, truncation is applied instead of RD. 
The contrasting behaviour is observed when $p_1=\pm1$, and $c=\pm(2^{-24}+2^{-25-i})$, with $i\ge 0$, for which the result is $d=1$ for the positive
and $d=-1-2^{-23}$ for the negative
case. 
The underlying reason is that part of the significand of $c$ remains within the first $25$-bit fractional bits of the product sum, allowing the round-down effect to occur. 

Finally, to determine whether odd- and even-indexed product sums are combined prior to adding $c$, we set $c=\pm(2^{-24}+2^{-25})$ and permute $\pm1$ and $\pm2^{-25}$ between $p_1$ and $p_2$, emulating odd- and even-indexed partial sums. 
This produces $d=1$ for positive values and $d=-(1+2^{-22})$ for negative values. This confirms that odd- and even-indexed product sums are first accumulated, followed by the addition of $c$. 
If any alternate ordering were present, at least one permutation would yield $d=1+2^{-23}$ for positive values. 
We obtained identical results for the cancellation test mentioned in Section~\ref{subsec_cdan3_fp16}. 
However, the odd- and even-indexed product terms separate accumulation requires a unique test in which one of the product sums is zero but its corresponding exponent is the maximum. 
For example, we set $c=2^{-20}$, $a_1=b_1=2^0$, and $a_2=-a_4=2^5$, $b_2=b_4=1$, which results in $d=1$ instead of $d=1+2^{-20}$. 
Although the even-indexed product sum is zero, the maximum exponent against which $c$ is aligned is $4$, rather than $0$.
Thus, the maximum exponent is used for alignment regardless of whether the corresponding significand sum is zero.

\subsubsection{CDNA3 summary}

\begin{figure}
    \centering
    \includegraphics{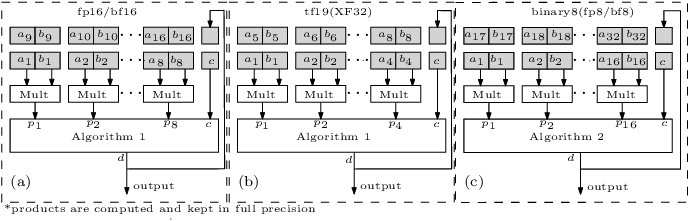}
    \caption{Matrix core model diagram in CDNA 3. (a) fp16 and bf16 input format, (b) tf19 input format, and (c) binary8 input format.}
    \label{fig:CNDA3}
\end{figure}

The CDNA~3 architecture's matrix core model is depicted in Fig.~\ref{fig:CNDA3}. The block diagram is supported by Algorithm~\ref{alg:cdna3_half} and Algorithm~\ref{alg:cdna3_fp8} for clarity. The binary32 input model corresponds to an SFMA and is therefore not shown; however, it is supported in the CDNA~3 architecture.

\begin{algorithm}[t]
\caption{CDNA 3 fp16/bf16/tf32 Input Formats Accumulation}
\label{alg:cdna3_half}
\begin{algorithmic}[1]
\Require{$p_i$ for $i=1,\dots,\nfma$, and $c$}
\Ensure{$d$}
\State
$e_{\mathrm{max}}=\mathrm{max}([e_{p_1},\dots,e_{p_1}])$
\State
    align $s_{p_i|1,\dots,\nfma}$ to $e_{\max}$, truncate to $24$ fractional bits, accumulate into $\psumsig$\;
\State $e_c=-126$ \textbf{ if } $c=0$
\If{$e_{\mathrm{max}}\ge e_{c}$}
\State $s_c'\gets$ shift $s_c$ to right by $(e_{\mathrm{max}}-e_c)$ and RD it to $24$ frac. bits
\State $\Sacc=\psumsig+ s_c'$\;
\Else
\State $\psumsigd\gets$ shift $S_{p_i,\mathrm{sum}}$ to right by $(e_c-e_{\mathrm{max}})$ and RD it to $32$ frac. bits.
\State $e_{\mathrm{max}}=e_c$;~$\Sacc=\psumsigd+ s_c$\;  
\State $\{\Sacc,e_{\mathrm{max}}\}\gets$ normalise $\{\Sacc,e_{\mathrm{exp}}\} ~\%$subnormal-aware normalisation\;
\State $\Sacc\gets$RD $\Sacc$ to 31 frac. bits
\EndIf
\State $d\gets$ convert $\{\Sacc,e_{\mathrm{max}}\}$ to fp32 with RNE
\end{algorithmic}
\end{algorithm}

\begin{algorithm}[t]
\caption{CDNA 3 Binary8 Input Format Accumulation}
\label{alg:cdna3_fp8}
\begin{algorithmic}[1]
\Require {$p_i$ for $i=1,\dots,\nfma$, and $c$}
\Ensure{$d$}

\State $e_{\mathrm{max-odd}}=\mathrm{max}([e_{p_1},e_{p_3},\dots,e_{p_\nfma-1}]),~e_{\mathrm{max-even}}=\mathrm{max}([e_{p_2},e_{p_4},\dots,e_{p_\nfma}])$\;
\State align odd and even indexed $s_{p_i}$  w.r.t $e_{\mathrm{max-odd}}$ and $e_{\mathrm{max-even}}$, respectively\; 
\State truncate to $24$ fractional bits and accumulate into $s_{p_i,\mathrm{sum-odd}},$ and $S_{p_i,\mathrm{sum-even}}$\;
\State
$e_{\max}=\max\{e_{\mathrm{max-odd}},e_\mathrm{max-even}\}$\;
\State
$\{S'_{p_i,\mathrm{sum-odd}},S'_{p_i,\mathrm{sum-even}}\}\gets$ right shift $\{S_{p_i,\mathrm{sum-odd}},S_{p_i,\mathrm{sum-even}}\}$ by $\{e_{\mathrm{max}}-e_{\mathrm{max-odd}},e_{\mathrm{max}}-e_\mathrm{max-even}\}$, respectively, and RD both to $24$ frac.\; 
\State
 $\psumsig=S'_{p_i,\mathrm{sum-odd}}+S'_{p_i,\mathrm{sum-even}}$\;
\State $e_c=-126$ \textbf{ if } $c=0$
\If{$e_{\mathrm{max}}\ge e_{c}$}
\If {$e_{\mathrm{max}}-e_c\le 25$}
\State 
$s_c'\gets$ shift $s_c$ to right by $(e_{\mathrm{max}}-e_c)$ and RD it to $24$ frac. bits
\Else
\State $s'_c=0$
\EndIf
\State $\Sacc=\psumsig+s'_c$
\Else
\State Follow step 8-11 of Algorithm~\ref{alg:cdna3_half}
\EndIf
\State $d\gets$ convert $\{\Sacc,e_{\mathrm{max}}\}$ to fp32 with RNE
\end{algorithmic}
\end{algorithm}

\section{Randomized Tests Suite and Refinement of Models}
\label{sec:EnsTest}

As our focus is on a single element of the output matrix $D$, the randomized testing procedure compares the result of (1) for independently generated vectors $a$, $b$, and scalar $c$ in each run.
While it makes sampling a lot easier, it does not make the sampling space small enough to be exhaustively searched.
Where there can be many ways to sample the input space, we propose following ones with appropriate justification.
With $N_{\mathrm{ens}}$ denoting the finite ensemble size, and $k$ the length of the input vectors $a$ and $b$ which denotes the inner dimension supported by an MFMA intrinsic instruction, we perform all testing and experiments using the maximum supported value of $k$.

In the verification process, we first generate a set of random inputs to evaluate the GPU matrix cores, with the outputs written to a text file. The same random inputs are then generated in MATLAB to evaluate the corresponding matrix-core models, followed by a comparison between the GPU and model outputs. We perform this verification on the MI100, MI210 and MI250
for fp16 and bf16 input formats, and on the MI300A/X (CDNA 3) for Binary8 (E5M2 and E4M3), fp16, bf16, and tf19. 
In total, we perform nine verification tests in two sampling strategies.

\subsection{Exponent-Mantissa Sampling}
\label{subsec_testing}

A straightforward approach would be to draw $a,b \sim \mathcal{U}$, where $\mathcal{U}$ denotes the uniform distribution. However, such sampling does not yield a uniform distribution of floating-point exponents. 
Rather, the exponent distribution is biased towards larger values.
To obtain a more uniform coverage of the floating-point exponent range, we generate the random vectors by sampling the significand uniformly in the range  $[-2+2^{-f+1},2-2^{-f+1}]$, where $f$ denotes the precision of the considered input format, as defined in Table~\ref{table:fp-formats}.
The corresponding exponents are drawn from a uniform integer distribution over the prescribed lower and upper bounds of the exponent range. 
The pseudo-code is outlined in Algorithm~\ref{alg:ensemble_genertion}.
With this sampling strategy, all architectures considered are tested on the basis of the input format exponent ranges described below. 
\sw{}{}
\begin{algorithm}
\caption{Ensemble Generation Psuedo-code for Randomized Testing of Tensor Core Models}
\label{alg:ensemble_genertion}
\begin{algorithmic}
\State Choose $k$ and $f$ based on the input format and GPU architecture.
\State mt19937 rng(0) \quad $\%~$Mersenne Twister with seed 0
\State choose $N_{\mathrm{ens}},~e_{\mathrm{a}}^{-},~e_{\mathrm{a}}^{+}, e_{\mathrm{b}}^{-},~e_{\mathrm{b}}^{+}, e^{-}_{\mathrm{c}},~e_{\mathrm{c}}^{+}$
\For{$i=1:N_{\mathrm{ens}}$}
\State $a_\ell = s_{\ell} 2^{e_\ell},\quad
e_\ell \overset{\mathrm{i.i.d.}}{\sim} \mathcal{U}(e_{\mathrm{a}}^{-},e_{\mathrm{a}}^{+}),\quad
s_{\ell} \overset{\mathrm{i.i.d.}}{\sim} \mathcal{U}(-2+2^{-f+1},\,2-2^{-f+1}),\quad
\ell=1,\ldots,k$
\State $b_\ell = s_{\ell} 2^{e_\ell},\quad
e_\ell \overset{\mathrm{i.i.d.}}{\sim} \mathcal{U}(e_{\mathrm{b}}^{-},e_{\mathrm{b}}^{+}),\quad
s_{\ell} \overset{\mathrm{i.i.d.}}{\sim} \mathcal{U}(-2+2^{-f+1},\,2-2^{-f+1}),\quad
\ell=1,\ldots,k.$
\State $c = s 2^{e},\quad
c \overset{\mathrm{i.i.d.}}{\sim} \mathcal{U}(e_{\mathrm{c}}^{-},e_{\mathrm{c}}^{+}),\quad
s \overset{\mathrm{i.i.d.}}{\sim} \mathcal{U}(-2+2^{-23},\,2-2^{-23})$\;
\EndFor
\end{algorithmic}
\end{algorithm}


\subsubsection{Half Precision (fp16)}
In this input format, all architectures are tested over $N_{\mathrm{est}}=10^6$ random inputs each with $k=16$ and following exponent limits:
\begin{itemize}
\item  $e_{\mathrm{a}}^{-}=e_{\mathrm{b}}^{-}=-14$, $e_{\mathrm{c}}^{-}=-126$, $e_{\mathrm{a}}^{+}=e_{\mathrm{b}}^{+}=15$, and $e_{\mathrm{c}}^{+}=127$. 
 \item $   e_{\mathrm{a}}^{-}=e_{\mathrm{b}}^{-}=-24$, $e_{\mathrm{c}}^{-}=-149$, $e_{\mathrm{a}}^{+}=e_{\mathrm{b}}^{+}=-14$, $e_{\mathrm{c}}^{+}=-126$. 
 This tests the models in subnormal range where the exponent limits are set between minimum normal and minimum subnormal exponents. 
 \end{itemize}

\subsubsection{BrainFloat16 (bf16)}
Analogously to the FP16 input-format case, we use $N_{\mathrm{ens}}=10^6$ and $k=16$ for CDNA 2 and 3, while for CDNA 1, $k$ is limited to a maximum value of $8$, with the exponent limits set as given below:
\begin{itemize}
\item     $e_{\mathrm{a}}^{-}=e_{\mathrm{b}}^{-}=e_{\mathrm{c}}^{-}=-126$, $e_{\mathrm{a}}^{+}=e_{\mathrm{b}}^{+}=63$, $e_{\mathrm{c}}^{+}=127$. 
The maximum exponent limit for $a$ and $b$ is set to $63$ to prevent the product from becoming infinite,
as CDNA 3 maps $|p| \geq 2^{128}$ to infinity. 
This setting is adopted to evaluate the model using normal inputs, while special cases such as infinities and NaNs are treated separately, as they do not provide meaningful practical validation.
 \item $   e_{\mathrm{a}}^{-}=e_{\mathrm{b}}^{-}=-133$, $e_{\mathrm{c}}^{-}=-149$, $e_{\mathrm{a}}^{+}=e_{\mathrm{b}}^{+}=e_{\mathrm{c}}^{+}=-126$ which tests the models between minimum normal and subnormal exponent limits for bf16 input format.
 \end{itemize}

\subsubsection{TensorFloat32 (tf19)}
This format is supported on CDNA 3 and subsequent CDNA architectures, and is denoted as XF32 in the MFMA intrinsic. 
With $k=8$ and $N_{\mathrm{ens}}=10^6$, the same exponent limits as for bf16 were used; however, the minimum subnormal exponent was set to $-136$.
We found zero mismatches.

\subsubsection{Binary8 (fp8-e5m2, fp8-e4m3)}
CDNA 3 supports this format as an input format for matrix core operations. 
The AMD intrinsic instructions for MFMA in 8-bit input formats, fp8-e5m2 and fp8-e4m3, are referred to as fp8 and bf8, respectively\footnote{\_\_builtin\_amdgcn\_mfma\_f32\_32x32x16\_fp8\_fp8, \_\_builtin\_amdgcn\_mfma\_f32\_32x32x16\_bf8\_bf8. 
Also support mixed, i.e.,\_fp8\_bf8}.
Unlike the OCP fp8 formats, these are implemented using the FNUZ encoding.
With $\Nes=10^{6}$ and $k=16$, following exponent limits are used.

\begin{itemize}
 \item fp8/bf8: $   e_{\mathrm{a}}^{-}=e_{\mathrm{b}}^{-}=-7/-15$, $e_{\mathrm{c}}^{-}=-126$, $e_{\mathrm{a}}^{+}=e_{\mathrm{b}}^{+}=7/15$, $e_{\mathrm{c}}^{+}=127$.
 \item fp8/bf8: $   e_{\mathrm{a}}^{-}=e_{\mathrm{b}}^{-}=-10/-17$, $e_{\mathrm{c}}^{-}=-149$, $e_{\mathrm{a}}^{+}=e_{\mathrm{b}}^{+}=-7/-15$, $e_{\mathrm{c}}^{+}=-126$.
 \end{itemize}
For normal values, we scan the exponent range from the minimum to the maximum normal exponent. For subnormal values, we scan the range from the minimum normal exponent down to the minimum subnormal exponent.
 Minimum normal exponent in FNUZ is different than OCP encoding as outlined in Table~\ref{table:fp-formats}.
 The CDNA 3 architecture is further evaluated at larger values of ($N_{\mathrm{ens}}$), with the underlying rationale discussed in the following subsections.
 
\subsection{Independent Bit  Sampling} 

This sampling strategy samples each bit independently from a Bernoulli distribution with equal probability. To avoid infinities in the input, which occur frequently for formats with small exponent ranges under this sampling strategy, we redraw the exponent whenever it results in an infinity. 

%
%
%
\subsection{Iterative Refinement Via Mismatch Based Hidden Feature Detection}

Across all configurations considered above, no mismatches were observed between the matrix-core model and the GPU outputs for CDNA 1 or CDNA 2, indicating that the feature space was sufficient for both sampling strategies. 
For CDNA 3, the independent bit-sampling strategy likewise produced no mismatches. 
However, the first sampling strategy revealed a small number of mismatches during model refinement, as touched upon in Section~\ref{subsec_cdan3_fp16}. 
These mismatches occurred at different stages of the refinement process as successive updates were introduced. 
Due to the two-stage RD operation and the associated intermediate normalisation, the model required four refinement iterations in total, as summarised in Table~\ref{tab:error_bounds}. The mismatches were progressively eliminated through these successive refinements. 
In the fourth refinement iteration, the validation was extended to $\Nes=10^7$ across the full exponent range using the first sampling strategy for fp16, bf16, and tf32, and to $2\times10^7$ for binary8. No mismatches were observed for any input format.
For $\Nes=10^6$, each test required, on average, $13\pm1.5$ minutes on the MATLAB side and $1$--$2$ minutes on the GPU.

\subsection{Probabilistic Analysis}
Suppose we are looking for triplets
$t=(a,b,c)$, where $a$,$b\in\mathbb{R}^k$, in which the result from the matrix core operation in hardware
differs from the one predicted by the software matrix core model. Suppose that the set
of such triplets causing such mismatches is $E$. Then, we would like
to estimate the probability $P(t\in E|S)$, that is, the probability,
given a single triplet $t$ sampled according to $S$, that $t$ produces
a model error. We will refer to this probability for short as $p_{S}$.

Let's suppose that we draw $\Nes$ samples, and observe $m$ errors.
The likelihood of this data is given by the binomial distribution:
\[
L(p_{S})=P_{Bin}(m|\Nes,p_{S})=\frac{\Nes!}{m!(\Nes-m)!}p^{m}_{S}(1-p_{S})^{\Nes-m}.
\]

{Assuming} a uniform prior distribution density on $p_{S}$, i.e. $\pi(p_{S})=1$,
$0\le p_{S}\le1${,} the posterior density for $p_{S}$ is
\begin{eqnarray*}
\pi(p_{S}|\Nes,m)= \kappa\times\pi(p_{S})\times L(p_{S}) = \frac{(\Nes+1)!}{m!(\Nes-m)!}p^{m}_{S}(1-p_{S})^{\Nes-m},
\end{eqnarray*}
{which is} a {Beta} distribution density on the {1}-simplex.
The ratio of factorials is a normalization {constant }guaranteeing that $\int^{1}_{0}dp_{S}\,\pi(p_{S}|\Nes,m)=1$.

{For} the case $m=0$, which is the usual case in which
sampling has not turned up any errors{, we have}
\[
\pi(p_{S}|\Nes,m=0)=(\Nes+1)\times(1-p_{S})^{\Nes}.
\]
This density peaks at $p_{S}=0$, falling off towards 0 at $p_{S}=1$.
We can construct an $\epsilon$-upper bound on $p_{S}$ denoted by $p_{\epsilon}$,
which is defined by the equation
\[
\int^{p_{\epsilon}}_{0}dp_{S}\,\pi(p_{S}|\Nes,m=0)=1-\epsilon.
\]
The quantity $1-\epsilon$ represents the level of certainty that
$p_{S}<p_{\epsilon}$. Thus, we may choose $\epsilon$ as small as
we like ($10^{-6}$, say), and solve for the $p_{\epsilon}$ that
produces this $\epsilon$.

The integral is
\begin{eqnarray*}
\int^{p_{\epsilon}}_{0}dp_{S}\,\pi(p_{S}|\Nes,m=0) & = & \int^{p_{\epsilon}}_{0}dp_{S}\,(\Nes+1)\times(1-p_{S})^{\Nes}\\
 & = & \int^{p_{\epsilon}}_{0}dp_{S}\,\frac{d}{dp_{s}}(1-p_{S})^{\Nes+1} = 1-(1-p_{\epsilon})^{\Nes+1}.
\end{eqnarray*}

Setting this equal to $1-\epsilon$ we obtain
\begin{eqnarray*}
p_{\epsilon} & = & 1-\epsilon^{\frac{1}{\Nes+1}}= 1-\exp\left(\frac{\log\epsilon}{\Nes+1}\right).
\end{eqnarray*}
In the limit of many samples, such that $\Nes+1\gg|\log\epsilon|$, this
becomes
$
p_{\epsilon}\approx\frac{\log\epsilon^{-1}}{\Nes+1}$.

We apply the probabilistic framework described above to quantify the
confidence bounds of our experimental validation.  
The number of tested samples, observed
mismatches, and the corresponding Bayesian upper bounds are summarized
in Table~\ref{tab:error_bounds} in each iteration of the refinement of the model.
Thus, for fixed $\epsilon$, a tenfold increase in error-free trials reduces the upper bound by approximately tenfold. For $\epsilon=10^{-6}$, $10^5$, $10^6$, and $10^7$ trials yield bounds of approximately $1.38\times10^{-4}$, $1.38\times10^{-5}$, and $1.38\times10^{-6}$, respectively. Hence, our choice of $10^6$--$10^7$ trials provides $10^{-5}$--$10^{-6}$-level bounds while avoiding the proportionally greater cost of validation on the Matlab side.

\begin{table}[t]
\centering
\caption{Experimental validation results across four model-refinement iterations and Bayesian upper bounds on the mismatch probability of the CDNA 3 matrix core model using exponent-mantissa sampling.}
\label{tab:error_bounds}
\begin{tabular}{c|c|c|c|c}
\hline
Iteration & Input Format & $\Nes$ & Mismatches$(m)$ &
Upper Bound on $p_{\epsilon=10^{-6}}$ \\
\hline
\multirow{4}{*}{0}
& binary8 (fp8, bf8) & $2\times10^{6}$ & 0 & $6.91\times10^{-6}$ \\
& fp16 & $10^{6}$ & 1 & $1.69\times10^{-5}$ \\
& bf16 & $10^{6}$ & 0 & $1.38\times10^{-5}$ \\
& tf32 & $10^{6}$ & 1 & $1.69\times10^{-5}$ \\
\hline
\multirow{4}{*}{1}
& binary8 (fp8, bf8) & $2\times10^{6}$ & 0 & $6.91\times10^{-6}$ \\
& fp16 & $10^{6}$ & 0 & $1.38\times10^{-5}$ \\
& bf16 & $10^{6}$ & 0 & $1.38\times10^{-5}$ \\
& tf32 & $10^{6}$ & 1 & $1.69\times10^{-5}$ \\
\hline
\multirow{4}{*}{2}
& binary8 (fp8, bf8) & $2\times10^{6}$ & 0 & $6.91\times10^{-6}$ \\
& fp16 & $10^{6}$ & 0 & $1.38\times10^{-5}$ \\
& bf16 & $10^{6}$ & 0 & $1.38\times10^{-5}$ \\
& tf32 & $10^{6}$ & 0 & $1.38\times10^{-5}$ \\
\hline
\multirow{1}{*}{3}
& fp16 & $1.25\times10^{6}$ & 1 & $1.35\times10^{-5}$\\ 
\hline
\multirow{4}{*}{4}
& binary8 (fp8, bf8) & $2\times10^{7}$ & 0 & $6.91\times10^{-7}$ \\
& fp16 & $10^{7}$ & 0 & $1.38\times10^{-6}$ \\
& bf16 & $10^{7}$ & 0 & $1.38\times10^{-6}$ \\
& tf32 & $10^{7}$ & 0 & $1.38\times10^{-6}$ \\
\hline
\end{tabular}
\end{table}

\section{Comparison of GPU Numerical Features}
While NVIDIA GPU numerical features have been reported in~\cite{fkmm_nvidia_tc,xie25_mmasim}, we briefly compare them with those of AMD GPUs in Table~\ref{tab:nfc}. 
The table summarizes key numerical features across several generations of NVIDIA tensor cores and AMD matrix cores, including product alignment, the FMA size, handling of the addend $c$, alignment and rounding of the accumulated product sum, and the minimum supported exponent. These features highlight the differences in accumulation, rounding, and precision handling across GPU MMA architectures.
The table speaks for itself---the approaches of evaluating dot products and matrix multiplies of floating-point numbers, in hardware, differ significantly between the two major companies.
This translates to numerical error accumulation differences, as shown in the demonstrative applications of Section~\ref{sec:appls}.

\begin{table}[htbp]
\centering
\scriptsize
\setlength{\tabcolsep}{2pt}
\renewcommand{\arraystretch}{1.1}

\caption{Comparison of numerical features of several generations of NVIDIA tensor cores (mixed-precision matrix multipliers)~\cite{fkmm_nvidia_tc} and AMD matrix cores. Newly derived models are in bold.}
\label{tab:nfc}

\resizebox{\textwidth}{!}{%
\begin{tabular}{lcccccccccc}
\toprule
& B200 & H100 & {\bf MI300A/X} & H100, B200 & A100 & {\bf MI100} &
{\bf MI300A/X} & H100, B200 & A100 & {\bf MI300A/X} \\
\midrule

In
& fp8 & fp8 & fp8
& fp16/bf16 & fp16/bf16 & fp16/bf16 & fp16/bf16
& tf19 & tf19 & tf19 \\

Prd. Align.
& (2,25) & (2,13) & (2,24)
& (2,25) & (2,24) & (1,$\infty$) & (2,24)
& (2,25) & (2,24) & (2,24) \\

$N_{\mathrm{FMA}}$
& 32 & 32 & $8_{\mathrm{odd-even}}$
& 16/16 & 8/8 & 4/2 & 8/8
& 8 & 4 & 4 \\

$e_{c=0}$
& $-\infty$ & $-\infty$ & $-126$
& $-\infty$ & $-\infty$ & $-\infty$ & $-126$
& $-\infty$ & $-\infty$ & $-126$ \\

$c$ late/early
& early & early & late
& early & early & early & late
& early & early & late \\

$\mathrm{align}(c, S_{p_i-\mathrm{sum}})$.
& - & - & (24,32,RD)
& - & - & - & (24,32,RD)
& - & - & (24,32,RD) \\

out. round.
& RTZ & RTZ & RNE
& RTZ & RTZ & RNE & RNE
& RTZ & RTZ & RNE \\

$e^{\mathrm{lim}}_{\min}$
& - & - & -
& -133 & -132 & - & -126
& -133 & -132 & -126 \\

$|p_i|$-overflow
& No & No & $2^{128}$
& No & No & No & $2^{128}$
& No & No & $2^{128}$ \\
\bottomrule
\end{tabular}%
    }

\vspace{2mm}

\begin{minipage}{0.95\textwidth}
\footnotesize
\textbf{Note:}
$(24,32,\mathrm{RD})$ indicates that shifted $s_c$ and $S_{p_i-\mathrm{sum}}$ are rounded to 24 and 32 frac. bits during alignment, respectively;
$\mathrm{odd\text{-}even}$ denotes the interleaved product-addition
pattern, where odd- and even-indexed products are accumulated separately. 
The feature $e_{c=0}$ denotes the exponent of $c$ when $c=0$, with $e_{c=0}=-\infty$ indicating that $c$ is discarded during accumulation.
MI210 and MI250 GPUs support fp16/bf16 and implement SFMA with the
grouping shown in Fig.~\ref{fig:CDNA2}.
The output is assumed to be in fp32 format. NVIDIA GPUs support fp16-format output for both fp8- and fp16-format inputs and implement $e_{\mathrm{min}}^{\mathrm{lim}}=\{-20,-21\}$ for A100 and H100/B200, respectively, across all GPUs, rather than $-14$, which is the minimum normal exponent of the fp16 format.
\end{minipage}

\end{table}
 \section{MFMA Software Models}

This section introduces MATLAB Matrix Core v0.6, comprising the extensions to the software developed as part of this research. The toolbox was developed in MATLAB R2026a and uses CPFloat~\cite{fami23} for conversion from double- or single-precision formats to all supported input formats. The toolbox is structured around the generalised matrix multiplier model, which supports four main architectural configurations: \texttt{correct\_rounding}, modelling a Kulisch-style correctly rounded accumulator such as CDNA 1; \texttt{pair\_wise\_sum}, modelling the pairwise accumulation used in CDNA 2; \texttt{global\_alignment}, modelling the CDNA 3-like architecture for fp16, bf16, and tf32 inputs; and \texttt{odd\_even\_grouping}, modelling the CDNA 3-like architecture for binary8 inputs. Each configuration provides parameters to control the relevant architectural features, including $\nfma$, alignment and rounding behaviour, subnormal support, and exponent handling, thereby allowing both the AMD and alternative configurations to be simulated. 
With the \texttt{Parallel Computing Toolbox} available, matrix multiplications using the proposed models are parallelized across available CPU cores. 
The toolbox is also compatible with GNU Octave and can be accessed from Python via Oct2Py or the MATLAB Engine API. In future work, we will port the model back-end to C to improve performance.
An example call of MI300A matrix core for a 4x4 matrix multiply accumulate is shown below.
\begin{tcolorbox}[colback=gray!10, colframe=gray!30,
top=0pt, bottom=0pt, left=3pt, right=0pt,   
    boxsep=0pt,]
\begin{lstlisting}
inopts.format = 'fp8-e4m3', outopts.format = 'binary32';
A = cpfloat(randn(4), inopts), B = cpfloat(randn(4), inopts);
C = cpfloat(randn(4), outopts), alpha = 1, beta = 1;
MI300AMC(alpha, A, B, beta, C, inopts.format, outopts.format)
>> ans =
    0.1484   -0.6631   -0.1836   -1.3271
    0.8232    1.6418    0.4805    3.0227
    3.6592   -0.1250    1.4902    2.2637
    3.7432   -3.4275    0.2031    0.1663
\end{lstlisting}
\end{tcolorbox}

\section{Application Scenarios}
\label{sec:appls}

In this section we apply the developed AMD matrix core models to several demonstration applications, acting as representative examples of how the models can be used.
The source code of these experiments is available.\footnote{\url{https://github.com/north-numerical-computing/MATLAB-tensor-core/tree/main/experiments}}

\subsection{Multi-word Arithmetic}

We have implemented an algorithm for emulating high-precision matrix multiplication via a multi-word representation of matrices in one of the low-precision formats supported by the AMD matrix cores.
The details are provided by Mary~and~Mikaitis~\cite{mami25}.
We have plotted three generations of the AMD CDNA GPUs and one of NVIDIA, for comparison.
In each diagram, the number of words determines the accuracy of the conversion of input matrices between binary64 representation and a $N$-word low-precision representation: larger $N$ means lower error in converting the binary64 inputs.

\begin{figure*}
  \begin{center}
    \footnotesize
    \begin{tikzpicture}
      \begin{groupplot}[
        group style={
          group size=3 by 3,
          vertical sep=1.2cm
        },
        ymode=log,
        xmode=log,
        width=1.9in,
        height=1.6in,
        grid=major,
        ymax = 10^(0),
        ymin = 10^(-9),
        every axis plot/.append style={very thick, mark repeat=3}
        ]

        \nextgroupplot[
        ylabel={$\frac{\norminf{\widehat{C}-C}}{\norminf{A}\norminf{B}}$},
        align=left,
        title={Single word (fp8-e5m2)},
        xlabel = {$n$}
        ]

        \addplot[color=blue!70, mark=x] table [x=n, y=error] {data/matmul_test_fp8-e5m2_binary32_words_1_model_mi300x.dat};
        \addplot[color=red!70, mark=o] table [x=n, y=error] {data/matmul_test_fp8-e5m2_binary32_words_1_model_b200.dat};

        \nextgroupplot[
        align=left,
        title={Quad-word (fp8-e5m2)},
        xlabel = {$n$}
        ]

        \addplot[color=blue!70, mark=x] table [x=n, y=error] {data/matmul_test_fp8-e5m2_binary32_words_4_model_mi300x.dat};
        \addplot[color=red!70, mark=o] table [x=n, y=error] {data/matmul_test_fp8-e5m2_binary32_words_4_model_b200.dat};

        \nextgroupplot[
        align=left,
        title={6-word (fp8-e5m2)},
        xlabel = {$n$}
        ]

        \addplot[color=blue!70, mark=x] table [x=n, y=error] {data/matmul_test_fp8-e5m2_binary32_words_6_model_mi300x.dat};
        \addplot[color=red!70, mark=o] table [x=n, y=error] {data/matmul_test_fp8-e5m2_binary32_words_6_model_b200.dat};

      \end{groupplot}
    \end{tikzpicture}

    \begin{tikzpicture}[trim axis left, trim axis right]
      \begin{axis}[
        title = {},
        legend columns=5,
        scale only axis,
        width=1mm,
        hide axis,
        /tikz/every even column/.append style={column sep=0.4cm},
        legend style={at={(0,0)},anchor=center,draw=none,
          legend cell align={left},cells={line width=0.75pt}},
        legend image post style={sharp plot},
        legend cell align={left},
        every axis plot/.append style={thick}
        ]
        \addplot [color=blue!70, mark=x] (0,0);
        \addplot [red!70, mark=o] (0,0);
        \legend{MI300X, B200};
      \end{axis}
    \end{tikzpicture}
  \end{center}
  \caption{Multi-word arithmetic experiment presented by Mary and Mikaitis~\cite[Sec.~5]{mami25} on the simulation of various tensor cores. We have reproduced the experiment on MI300X and NVIDIA B200 matrix multipliers modelled in MATLAB. Relative norm-wise errors of matrix multiplication, compared with a default MATLAB binary64 multiplication, are shown. The input matrices to the GEMM are $A \in R^{10\times n}$ and $B \in R^{n \times 10}$. These matrices are multiplied with a multi-word algorithm \cite[Sec.~4]{mami25} by splitting them into several fp8-e5m2 words.}
   \label{fig:multi-word0}
\end{figure*}
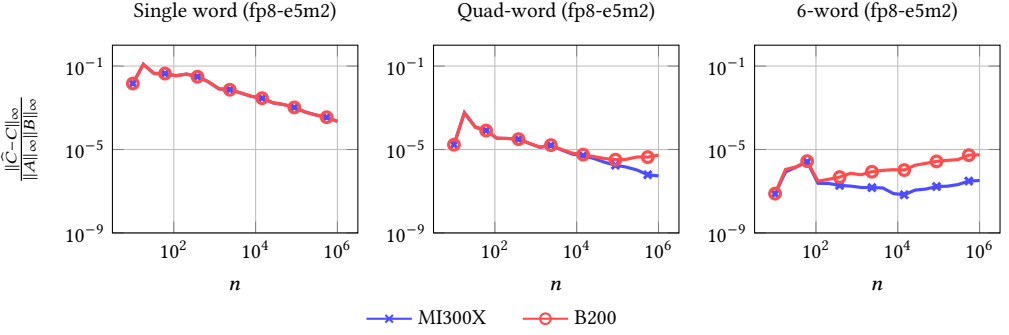

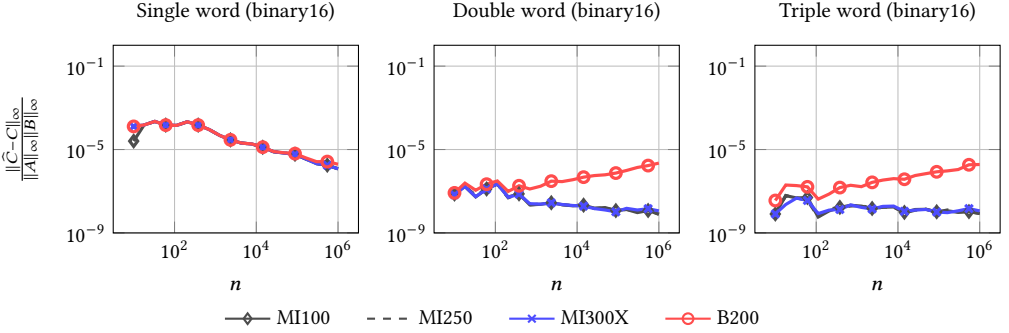
\begin{figure*}
  \begin{center}
    \footnotesize
    \begin{tikzpicture}
      \begin{groupplot}[
        group style={
          group size=3 by 3,
          vertical sep=1.2cm
        },
        ymode=log,
        xmode=log,
        width=1.9in,
        height=1.6in,
        grid=major,
        ymax = 10^(0),
        ymin = 10^(-9),
        every axis plot/.append style={very thick, mark repeat=3}
        ]

        \nextgroupplot[
        ylabel={$\frac{\norminf{\widehat{C}-C}}{\norminf{A}\norminf{B}}$},
        align=left,
        title={Single word (binary16)},
        xlabel = {$n$}
        ]

        \addplot[color=black!70, mark=diamond] table [x=n, y=error] {data/matmul_test_binary16_binary32_words_1_model_mi100.dat};
        \addplot[color=black!70, dashed] table [x=n, y=error] {data/matmul_test_binary16_binary32_words_1_model_mi250.dat};
        \addplot[color=blue!70, mark=x] table [x=n, y=error] {data/matmul_test_binary16_binary32_words_1_model_mi300x.dat};
        \addplot[color=red!70, mark=o] table [x=n, y=error] {data/matmul_test_binary16_binary32_words_1_model_b200.dat};

        \nextgroupplot[
        align=left,
        title={Double word (binary16)},
        xlabel = {$n$}
        ]

        \addplot[color=black!70, mark=diamond] table [x=n, y=error] {data/matmul_test_binary16_binary32_words_2_model_mi100.dat};
        \addplot[color=black!70, dashed] table [x=n, y=error] {data/matmul_test_binary16_binary32_words_2_model_mi250.dat};
        \addplot[color=blue!70, mark=x] table [x=n, y=error] {data/matmul_test_binary16_binary32_words_2_model_mi300x.dat};
        \addplot[color=red!70, mark=o] table [x=n, y=error] {data/matmul_test_binary16_binary32_words_2_model_b200.dat};

        \nextgroupplot[
        align=left,
        title={Triple word (binary16)},
        xlabel = {$n$}
        ]

        \addplot[color=black!70, mark=diamond] table [x=n, y=error] {data/matmul_test_binary16_binary32_words_3_model_mi100.dat};
        \addplot[color=black!70, dashed] table [x=n, y=error] {data/matmul_test_binary16_binary32_words_3_model_mi250.dat};
        \addplot[color=blue!70, mark=x] table [x=n, y=error] {data/matmul_test_binary16_binary32_words_3_model_mi300x.dat};
        \addplot[color=red!70, mark=o] table [x=n, y=error] {data/matmul_test_binary16_binary32_words_3_model_b200.dat};

      \end{groupplot}
    \end{tikzpicture}

    \begin{tikzpicture}[trim axis left, trim axis right]
      \begin{axis}[
        title = {},
        legend columns=5,
        scale only axis,
        width=1mm,
        hide axis,
        /tikz/every even column/.append style={column sep=0.4cm},
        legend style={at={(0,0)},anchor=center,draw=none,
          legend cell align={left},cells={line width=0.75pt}},
        legend image post style={sharp plot},
        legend cell align={left},
        every axis plot/.append style={thick}
        ]
        \addplot [black!70, mark=diamond] (0,0);
        \addplot [black!70, dashed] (0,0);
        \addplot [color=blue!70, mark=x] (0,0);
        \addplot [red!70, mark=o] (0,0);
        \legend{MI100, MI250, MI300X, B200};
      \end{axis}
    \end{tikzpicture}
  \end{center}
  \caption{Multi-word arithmetic experiment presented by Mary and Mikaitis~\cite[Sec.~5]{mami25} on the simulation of various tensor cores. Relative norm-wise errors of matrix multiplication, compared with a default MATLAB binary64 multiplication, are shown. The input matrices to the GEMM are $A \in R^{10\times n}$ and $B \in R^{n \times 10}$. These matrices are multiplied with a multi-word algorithm \cite[Sec.~4]{mami25} by splitting them into several fp16 words.}
   \label{fig:multi-word1}
\end{figure*}

Figure~\ref{fig:multi-word0} shows the experiments with fp8-e5m2.
The plot shows the relative norm-wise error as the inner dimension $n$ of two matrices being multiplied is increased.
The main differences between the MI300X and the B200 appear when either the dimension $n$ is relatively large or when six words are used.
This can be explained by the differences between these two devices as shown in Table~\ref{tab:nfc}.
Figures~\ref{fig:multi-word1}~and~\ref{fig:multi-word2} show similar experiments for fp16 and bf16.
With these formats, CDNA 1 and 2 can also be utilised.
Overall, all CDNA matrix multipliers yield similar error in this matrix multiplication algorithm, whilst B200 consistently has a larger error which grows with $n$. 
This behavior can be attributed to the fact that NVIDIA tensor cores use RTZ, whereas AMD matrix cores use RNE for the final rounding.
In addition, AMD applies RD when aligning the accumulation term with the product sum. However, the effect of this asymmetric rounding is not pronounced because the accumulation term often exceeds the inner-product term. Consequently, RD becomes effective only beyond the 31st fractional bit, which provides sufficient precision to reduce the impact of its asymmetric behavior on the final result.

\begin{figure*}
  \begin{center}
    \footnotesize
    \begin{tikzpicture}
      \begin{groupplot}[
        group style={
          group size=3 by 3,
          vertical sep=1.2cm
        },
        ymode=log,
        xmode=log,
        width=1.9in,
        height=1.6in,
        grid=major,
        ymax = 10^(0),
        ymin = 10^(-9),
        every axis plot/.append style={very thick, mark repeat=3}
        ]

        \nextgroupplot[
        ylabel={$\frac{\norminf{\widehat{C}-C}}{\norminf{A}\norminf{B}}$},
        align=left,
        title={Single word (bfloat16)},
        xlabel = {$n$}
        ]

        \addplot[color=black!70, mark=diamond] table [x=n, y=error] {data/matmul_test_bfloat16_binary32_words_1_model_mi100.dat};
        \addplot[color=black!70, dashed] table [x=n, y=error] {data/matmul_test_bfloat16_binary32_words_1_model_mi250.dat};
        \addplot[color=blue!70, mark=x] table [x=n, y=error] {data/matmul_test_bfloat16_binary32_words_1_model_mi300x.dat};
        \addplot[color=red!70, mark=o] table [x=n, y=error] {data/matmul_test_bfloat16_binary32_words_1_model_b200.dat};

        \nextgroupplot[
        align=left,
        title={Double word (bfloat16)},
        xlabel = {$n$}
        ]

        \addplot[color=black!70, mark=diamond] table [x=n, y=error] {data/matmul_test_bfloat16_binary32_words_2_model_mi100.dat};
        \addplot[color=black!70, dashed] table [x=n, y=error] {data/matmul_test_bfloat16_binary32_words_2_model_mi250.dat};
        \addplot[color=blue!70, mark=x] table [x=n, y=error] {data/matmul_test_bfloat16_binary32_words_2_model_mi300x.dat};
        \addplot[color=red!70, mark=o] table [x=n, y=error] {data/matmul_test_bfloat16_binary32_words_2_model_b200.dat};

        \nextgroupplot[
        align=left,
        title={Triple word (bfloat16)},
        xlabel = {$n$}
        ]

        \addplot[color=black!70, mark=diamond] table [x=n, y=error] {data/matmul_test_bfloat16_binary32_words_3_model_mi100.dat};
        \addplot[color=black!70, dashed] table [x=n, y=error] {data/matmul_test_bfloat16_binary32_words_3_model_mi250.dat};
        \addplot[color=blue!70, mark=x] table [x=n, y=error] {data/matmul_test_bfloat16_binary32_words_3_model_mi300x.dat};
        \addplot[color=red!70, mark=o] table [x=n, y=error] {data/matmul_test_bfloat16_binary32_words_3_model_b200.dat};

      \end{groupplot}
    \end{tikzpicture}

    \begin{tikzpicture}[trim axis left, trim axis right]
      \begin{axis}[
        title = {},
        legend columns=5,
        scale only axis,
        width=1mm,
        hide axis,
        /tikz/every even column/.append style={column sep=0.4cm},
        legend style={at={(0,0)},anchor=center,draw=none,
          legend cell align={left},cells={line width=0.75pt}},
        legend image post style={sharp plot},
        legend cell align={left},
        every axis plot/.append style={thick}
        ]
        \addplot [black!70, mark=diamond] (0,0);
        \addplot [black!70, dashed] (0,0);
        \addplot [color=blue!70, mark=x] (0,0);
        \addplot [red!70, mark=o] (0,0);
        \legend{MI100, MI250, MI300X, B200};
      \end{axis}
    \end{tikzpicture}
  \end{center}
  \caption{Multi-word arithmetic experiment presented by Mary and Mikaitis~\cite[Sec.~5]{mami25} on the simulation of various tensor cores. Relative norm-wise errors of matrix multiplication, compared with a default MATLAB binary64 multiplication, are shown. The input matrices to the GEMM are $A \in R^{10\times n}$ and $B \in R^{n \times 10}$. These matrices are multiplied with a multi-word algorithm \cite[Sec.~4]{mami25} by splitting them into several bffloat16 words.}
   \label{fig:multi-word2}
\end{figure*}

\subsection{Broadband Signal Processing Algorithms}

Many signal processing algorithms rely heavily on GEMMs, including applications in tensor decomposition, polynomial matrix algebra, and multiple-input multiple-output (MIMO) systems. Here, we investigate the impact of reduced-precision GEMM implementations on the output of a sequential matrix diagonalisation (SMD)~\cite{SMD} algorithm, a polynomial eigenvalue decomposition (PEVD) method used in broadband array processing applications~\cite{PEVD_app_25,PEVD_ISPM}. 

We consider a ${\ive{R}(z)}=\sum_{\tau}\ve{R}[\tau]z^{-\tau}\in\mathbb{R}^{8\times 8}$, where $\tau\in\mathbb{Z}$, with polynomial order $8$ as an example case. 
The off-diagonal energy threshold $\epsilon$ and the truncation parameter $\mu$ are both set $10^{-4}$ with algorithm allowed to iterate for $250$ iterations. 
The reference result is obtained from a MATLAB double-precision implementation of the complete SMD algorithm. 
For the GEMM-based implementations, only the computationally dominant GEMM operation, i.e., updating the paraunitary and partially diagonalized matrices, is replaced by bit-accurate tensor and matrix core models representing different GPU architectures.

\begin{figure}[htbp]
\centering
\begin{tikzpicture}
    \node {\includegraphics[width=\linewidth]{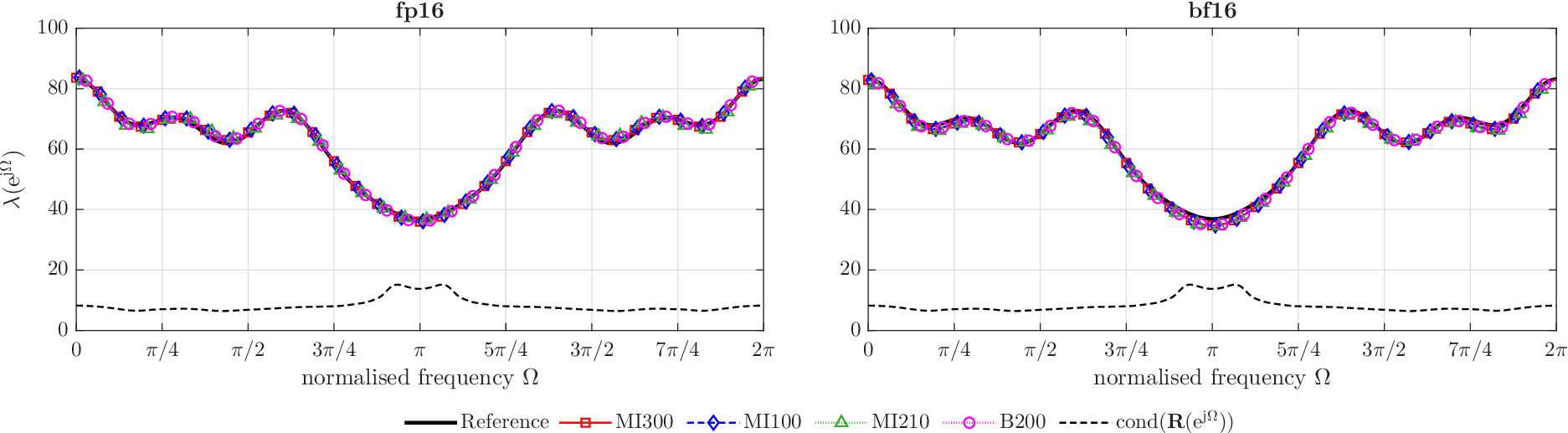}};
\end{tikzpicture}
\caption{Dominant eigenvalue of the example parahermitian polynomial matrix estimated using the SMD algorithm. The MATLAB double-precision GEMM implementation is used as the reference, while the remaining curves correspond to GEMM computations performed on different GPU matrix and tensor core architectures.}
\label{fig:pevd}
\end{figure}

\begin{figure}[htbp]
\centering
\begin{tikzpicture}
    \node {\includegraphics[width=\linewidth]{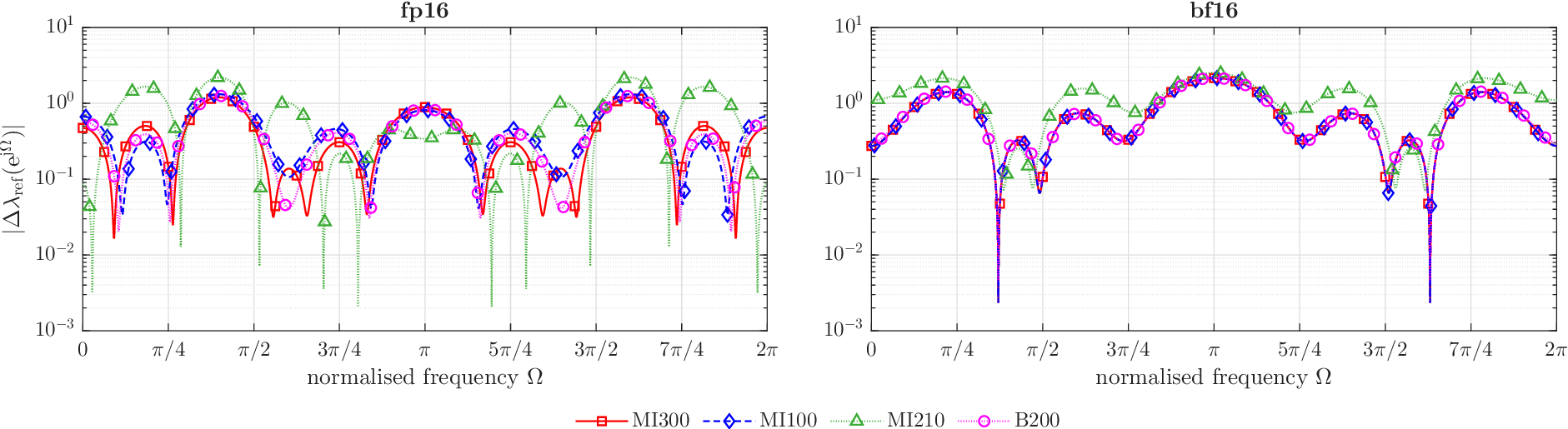}};
\end{tikzpicture}
\caption{Absolute difference in the estimated dominant eigenvalue of the example parahermitian matrix between the reference model (in double precision), and the various matrix and tensor-core models in fp16 and bf16 input formats.}
\label{fig:pevd_diff_ref}
\end{figure}

\begin{figure}[htbp]
\centering
\begin{tikzpicture}
    \node {\includegraphics[width=\linewidth]{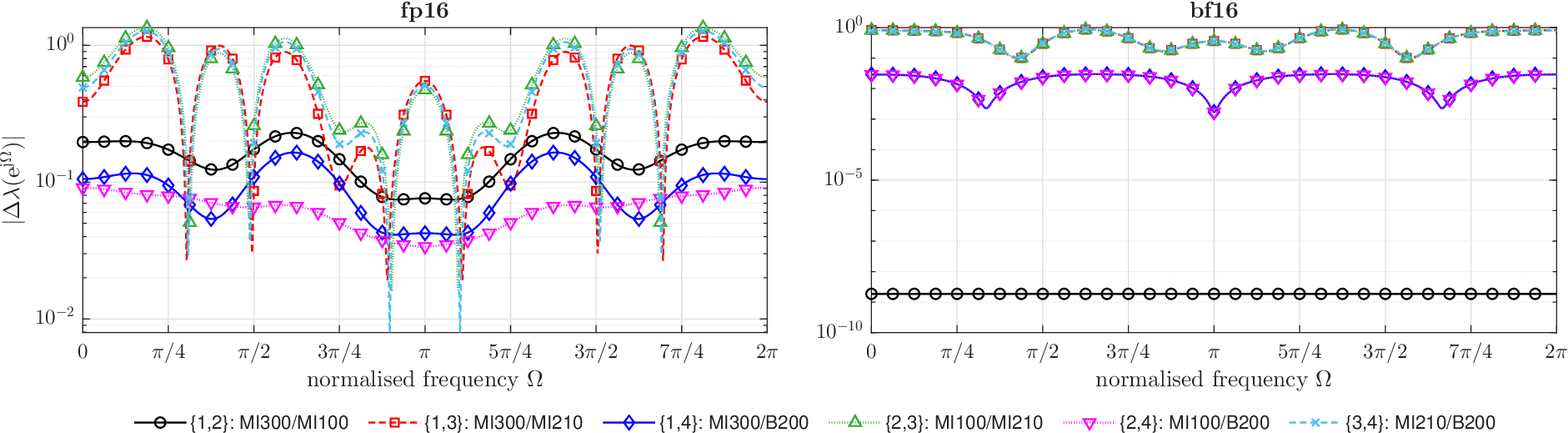}};
\end{tikzpicture}
\caption{Magnitude of the pairwise difference in the dominant eigenvalue estimated using the SMD algorithm due to GEMM computations on different GPU tensor or matrix core models. Each curve represents
$|\Delta\lambda_{i,j}(\mathrm{e}^{\mathrm{j}\Omega})|
= |\lambda_i(\mathrm{e}^{\mathrm{j}\Omega})-\lambda_j(\mathrm{e}^{\mathrm{j}\Omega})|$,
where $i$ and $j$ denote the GPU pair identified in the legend.}
\label{fig:pevd_diff}
\end{figure}

In this demonstration, in addition to CDNA 1, 2, and 3 architectures, we also consider the B200 NVIDIA GPU for both fp16 and bf16 input formats. The dominant eigenvalue and the condition number of the example polynomial matrix are shown in Figure~\ref{fig:pevd}. Although no visible difference can be observed between the eigenvalue estimates obtained using different GEMM models, the difference is non-zero, as illustrated in Figure~\ref{fig:pevd_diff_ref} and ~\ref{fig:pevd_diff}. 
The frequency-dependent variations i.e., the dip around certain frequencies, are difficult to attribute to any particular feature  because the error propagation depends on the repeated perturbation of the intermediate matrix $\ive{D}(z)$ during subsequent SMD iterations. 
These perturbations may either accumulate towards or away from the reference solution at different frequencies, and therefore, require a separate analysis.

The absolute difference between the GEMM models, presented in Fig.~\ref{fig:pevd_diff}, we can see that MI210 architecture exhibits comparatively larger differences from the other models, which may be attributed to its sequential fused multiply-add (SFMA) implementation and lack of subnormal number support. For fp16 inputs, the B200 shows the smallest difference with respect to MI100. This behaviour may be related to the two additional alignment bits available in B200, which provide behaviour closer to correctly rounded arithmetic compared with the other architectures.

For bf16 inputs, the MI300 and MI100 models exhibit almost identical behaviour. 
A possible explanation is that lower product significand precision reduces the difference between the finite and full-precision multi-term adders. 
This is further supported by the use of RNE rounding in both architectures. 
For example, bf16 products have lower significand precision, making them less susceptible to differences introduced by the 24-fractional-bit accumulator used by CDNA 3. 
In contrast, fp16 products have 22-bit significand precision and are therefore more susceptible to such differences.
Since the observed difference is small but not exactly zero, this suggests the presence of rare out-of-range alignment cases.
 
It should be noted that these results are not conclusive. 
For a different polynomial matrix, the observed behaviour may differ because the iterative nature of the SMD algorithm causes numerical perturbations to be repeatedly propagated through subsequent iterations. These perturbations may lead the algorithm either closer to or further from the reference output at each frequency. Nevertheless, this example demonstrates that small numerical differences introduced by reduced-precision GEMM implementations can influence the final result of iterative matrix algorithms. In subspace-based applications~\cite{MUSIC}, such deviations may manifest as subspace leakage, where errors in the estimated eigenvectors or invariant subspaces degrade the separation between signal and noise components, thereby deteriorating the performance. 
In NN-based polynomial eigenvalue extraction~\cite{PEVD_NN}, these differences may be more pronounced if both the SMD algorithm and NN training and inference are performed on the matrix multiplier across different GPU architectures.
Experiments with fp8 precision were also conducted; however, the estimated dominant eigenvalues exhibited significantly larger deviations from the double-precision reference. Therefore, fp8 precision is considered unsuitable for the considered example case.

This example demonstrates that seemingly minor numerical differences in low precision GEMM implementations can propagate through iterative matrix algorithms and influence the final eigenvalue, and therefore the corresponding eigenvector, estimates. Therefore, numerical characteristics of tensor and matrix core architectures must be carefully considered when deploying reduced-precision GEMM operations in signal processing applications.   

\section{Conclusion}
Matrix multipliers in AMD CDNA 1, CDNA 2, and CDNA 3 architectures support a wide range of low-precision input formats; however, their internal numerical behaviors contain diverse architectural features that are not publicly documented. 
While prior work has characterized these features~\cite{xie25_mmasim}, we systematically re-determine all features of CDNA 1, 2, and 3. Our results match those reported in the GitHub repository but differ from those reported in~\cite{xie25_mmasim}. 
We also provide the test vectors used to identify the numerical features, which can be used to characterize future architectures. 
Additionally, we provide flexible software models of the CDNA 1, 2, and 3 matrix cores that allow users to modify individual numerical features for numerical analysis. 
The models are validated in MATLAB against GPU results using two sampling strategies, with randomized test suites exceeding $10^7$ samples for all supported input formats and associated probabilistic upper bounds on the mismatch probability. The models also work in GNU Octave and are accessible within Python-based workflows through available language interfaces.

Our testing analysis reveals that the numerical features of AMD CDNA matrix cores vary significantly across generations and differ fundamentally from previously reported behaviors of NVIDIA GPU tensor cores. Specifically, CDNA 1 employs full-precision accumulation, CDNA 2 implements sequential FMA-based pair-wise accumulation approach, while CDNA 3 introduces a distinct numerical behavior that differs from both earlier AMD architectures and NVIDIA implementations. 

Finally, we demonstrate the practical impact of these numerical differences through two application scenarios: multi-word arithmetic and polynomial matrix algebra. In both cases, small variations in matrix multiplication behavior can accumulate over repeated operations and lead to different computational outcomes. These examples emphasize the importance of understanding the numerical characteristics of hardware accelerators when developing applications.

\section{Acknowledgment}
    We thank The COSmology MAchine (COSMA) support team at Durham University for providing the access to MI100/210/300A/300X.
    We thank Argonne
National Laboratory for providing access to MI100/250/300 GPUs.
FK and MM are funded by the EPSRC grant ``\emph{Informing Future Numerical Standards by Determining Features of Non-Standard Mathematical Hardware}'', ref. UKRI151.
CG was supported by the Office of Advanced Scientific Computing Research, Office of Science, U.S. Department of Energy, under Contract DE-AC02-06CH11357.


\bibliographystyle{ACM-Reference-Format}
\bibliography{references}

\end{document}